\documentclass[aps,reprint,amsmath,amssymb,prfluids,floatfix,onecolumn,superscriptaddress]{revtex4-2}

\usepackage{graphicx}
\usepackage{dcolumn}
\usepackage{bm}

\begin{document}

\preprint{AIP/123-QED}

\title[Sample title]{A Voronoi-tessellation-based approach for detection of coherent structures in sparsely-seeded flows}


\author{F. A. C. Martins}
\author{D. E. Rival}%
\email{d.e.rival@queensu.ca}
\affiliation{Department of Mechanical and Materials Engineering, Queen's University, Kingston, ON, Canada}

\date{\today}


\begin{abstract}
A novel algorithm to detect coherent structures with sparse Lagrangian particle tracking data, using Voronoi tessellation and techniques from spectral graph theory, is tested. Neighbouring tracer particles are naturally identified through the Voronoi tessellation of the tracers' distribution. The method examines the \textit{neighbouring time} of tracer trajectories, defined as the total flow time two Voronoi cells share a common Voronoi edge, by converting this information into a Cartesian distance in the graph representation of the Voronoi diagram. Coherence is assigned to groups of Voronoi cells whose neighbouring time remains high throughout the time interval of analysis. The technique is first tested on the two-dimensional synthetic data of a double-gyre flow, and then with challenging, large-scale three-dimensional Lagrangian particle tracking data behind a bluff body at high Reynolds number. The tested technique proves to be successful at identifying coherence with realistic experimental data. Specifically, it is shown that coherent tracer motion is identifiable for mean inter-particle distances of the order of the largest length scales in the flow.
\end{abstract}


\maketitle

\section{Introduction}
\label{sec:Introduction}

There are numerous concepts that describe the notion of coherent behaviour in time-dependent flows \citep{peacock2010introduction, hadjighasem2017critical}. The Eulerian framework for coherent structure detection typically consists of mapping Lagrangian particle tracking (LPT) data onto a grid to obtain a region in space in which computations of the velocity gradient tensor take place. The concept of Lagrangian coherent structures has also been proposed to identify material surfaces in a velocity field that extremize a certain stretching or shearing quantity \citep{haller2000lagrangian}. Compact groups of trajectories in flows have also been identified through clustering techniques \citep{froyland2015rough}, which revealed to be a useful tool even for incomplete trajectory data. Nonetheless, for a flow with largest length scales on the order of $D$, the array of current coherent-structures descriptions listed above (see algo \citep{banisch2017understanding, padberg2017network}) requires access to LPT data of mean inter-particle distances of $\sim \mathcal{O}(D \times 10^{-2})$ or less, which is not readily available in many \textit{in situ} engineering and biological flows \citep{schmale2015highways,shuckburgh2009robustness,davis1991observing,fujii2019observing}. Hence, the need for techniques to analyse sparsely-distributed track data persists. In this study, we discuss an alternative concept for coherent-structure detection based on Voronoi tessellation and graph clustering that has potential to detect coherent motion in realistic, sparsely-seeded fluid flows.

To avoid the inherent challenges associated with experimentally-acquired LPT data (e.g., sparse and inhomogeneous track sending), new approaches based on spectral clustering techniques have been assessed. Among these, one of the most promising techniques for applications with realistic, naturally-occorruing particle distribuitions is the Coherent-Structure Colouring (CSC) approach, originally tested by \citet{schlueter2017coherent}, and critically assessed by \citet{schlueter2017identification}, \citet{schlueter2019model}, and \citet{husic2019simultaneous}. The CSC algorithm adopts spectral graph clustering theory to colour-code LPT data based on the kinematic dissimilarity between all pairs of particles. The major advantage of such approach if the fact that the number of clusters (coherent structures) present in the flow does not need to be determined \textit{a priori}, avoiding the need of biased inputs from the user, such as needed in classical clustering approaches, e.g., $K$-means and fuzzy clustering, widely adopted in the literature \citep{saqib2017detecting,zhao2020k,hadjighasem2016spectral}.  Nonetheless, the CSC technique presents limitations when adopted to data sets from real-life applications: the CSC approach requires data-sets comprised of tracks of equal length, which is infeasible in most experimental settings. Moreover, since particles in the CSC approach are viewed as scattered, discrete points in space, the vast majority of the physical portion of the fluid domain is left empty, making diagnostics difficult to interpret for sparse seeding distributions. Nonetheless, as for the array of techniques listed above, the CSC approach was also found to have limited value for mean inter-particle distances of $\sim 0.15D$ or greater \citep{martins2021detection}, which restricts the analyses of coherent structures in challenging applications \citep{dimarco2005statistical}, motivating the current study.

\begin{figure}
  \centering
  \def\svgwidth{\columnwidth}
\begingroup%
  \makeatletter%
  \providecommand\color[2][]{%
    \errmessage{(Inkscape) Color is used for the text in Inkscape, but the package 'color.sty' is not loaded}%
    \renewcommand\color[2][]{}%
  }%
  \providecommand\transparent[1]{%
    \errmessage{(Inkscape) Transparency is used (non-zero) for the text in Inkscape, but the package 'transparent.sty' is not loaded}%
    \renewcommand\transparent[1]{}%
  }%
  \providecommand\rotatebox[2]{#2}%
  \newcommand*\fsize{\dimexpr\f@size pt\relax}%
  \newcommand*\lineheight[1]{\fontsize{\fsize}{#1\fsize}\selectfont}%
  \ifx\svgwidth\undefined%
    \setlength{\unitlength}{375.21645425bp}%
    \ifx\svgscale\undefined%
      \relax%
    \else%
      \setlength{\unitlength}{\unitlength * \real{\svgscale}}%
    \fi%
  \else%
    \setlength{\unitlength}{\svgwidth}%
  \fi%
  \global\let\svgwidth\undefined%
  \global\let\svgscale\undefined%
  \makeatother%
  \begin{picture}(1,0.61044621)%
    \lineheight{1}%
    \setlength\tabcolsep{0pt}%
    \put(0,0){\includegraphics[width=\unitlength,page=1]{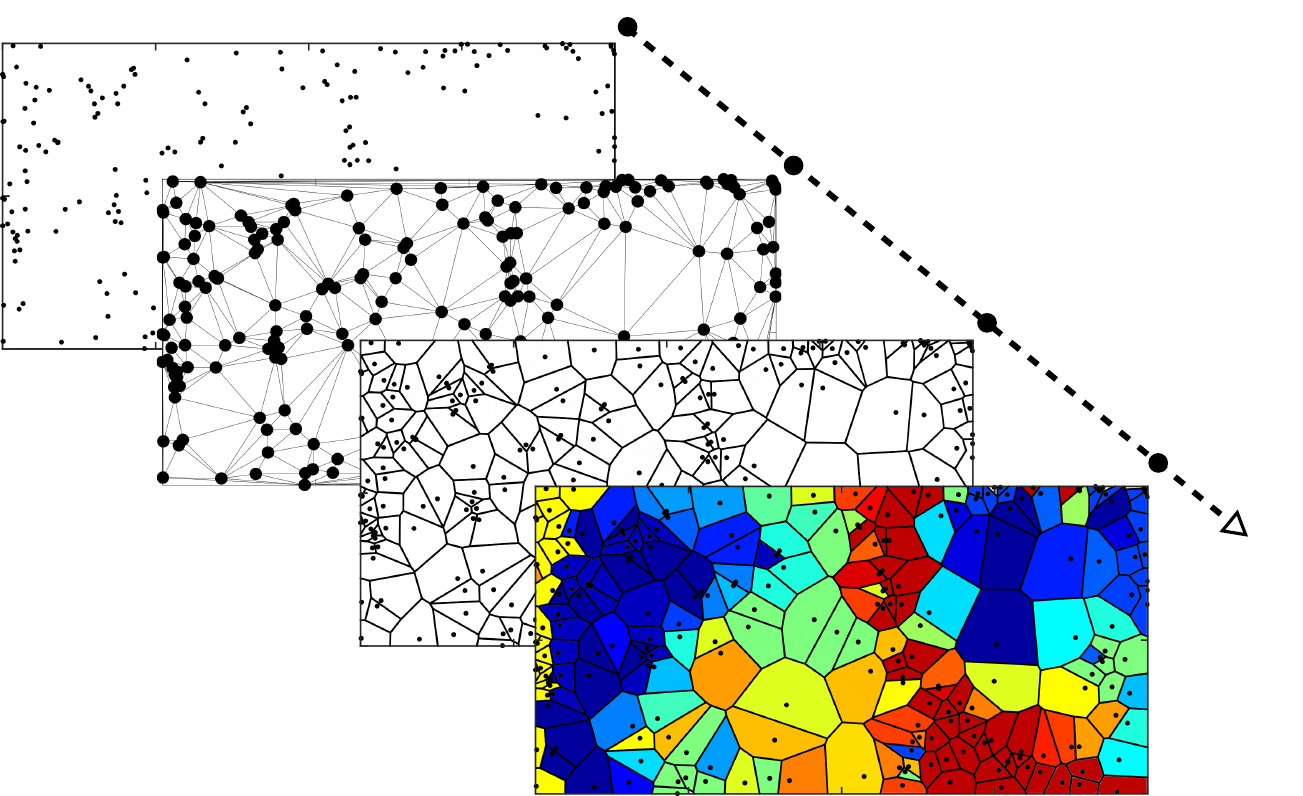}}%
    \put(0.4903408,0.60180193){\makebox(0,0)[lt]{\lineheight{1.25}\smash{\begin{tabular}[t]{l}Seeding particles\end{tabular}}}}%
    \put(0.76520863,0.36900118){\makebox(0,0)[lt]{\lineheight{1.25}\smash{\begin{tabular}[t]{l}Voronoi tessellation\end{tabular}}}}%
    \put(0.62230737,0.49253474){\makebox(0,0)[lt]{\lineheight{1.25}\smash{\begin{tabular}[t]{l}Delaunay triangulation\end{tabular}}}}%
    \put(0.89631136,0.28954746){\makebox(0,0)[lt]{\lineheight{1.25}\smash{\begin{tabular}[t]{l}Clustered \\Voronoi cells\end{tabular}}}}%
  \end{picture}%
\endgroup%
  
  \caption{Schematic of coherent-structure detection based on Voronoi tessellation and graph clustering. Voronoi diagram is built from the Delaunay triangulation of the seeding particles. Neighbouring Voronoi cells are then identified through the Delaunay triangulation of the tracer distribution. Neighbouring time of Voronoi cells acts as a higher-dimensional criterion for particle clustering, and coherent structures are identified as the clusters of Voronoi cells that persist for longer periods of time as neighbouring cells.}
  \label{fig:Schematics}
\end{figure}
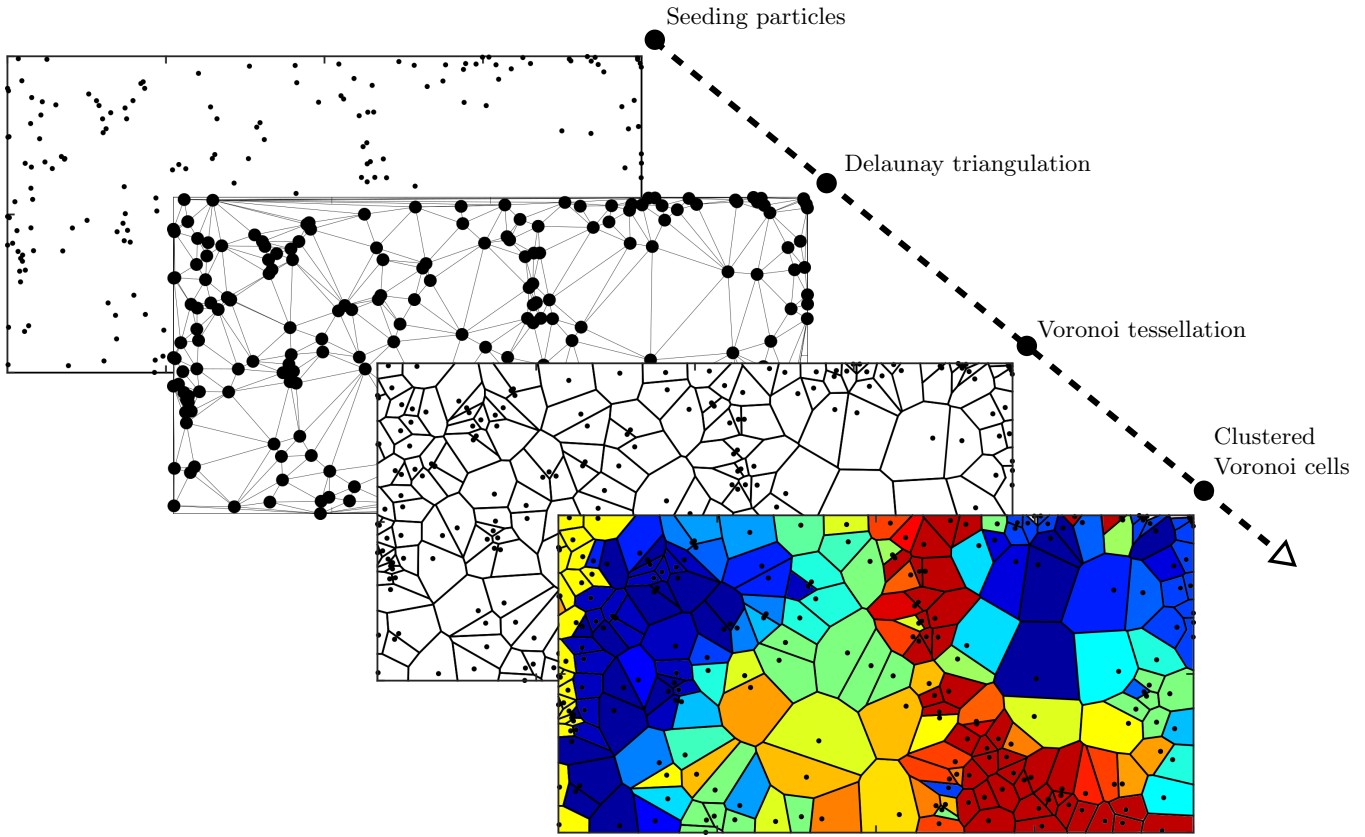

In the present study, rather than viewing a fluid parcel as a means for discretization, we employ the simple concept that a seeding particle, with its associated position and momentum, is representative of a volume of the fluid flow \citep{espanol2009voronoi, rosi2015lagrangian, padberg2017network, krueger2019quantitative}. One of the most natural ways of assigning volume to a tracer is through a Voronoi tessellation \citep{neeteson2015pressure}. The technique proposed in this study employs a Voronoi tessellation to detect neighbouring particles in the flow, using the period of time a pair of Voronoi cells persist as neighbouring cells, referred to here as their \textit{neighbouring time}, as a metric for coherence. The algorithm then employs spectral graph clustering theory to compare the neighbouring times of an arbitrary number of Voronoi cells, colour-coding them based on the period of time they share a common Voronoi edge (as shown in Fig. \ref{fig:Schematics}). In practice, Voronoi cells that present relatively long neighbouring times are colour-coded with similar colours. Since the neighbouring time is independent of reference frame, the tesded technique is naturally objective. Moreover, like the CSC technique, the current approach is based on spectral clustering, which does not necessarily constrain the technique to a minimum number of data points, making it a strong candidate for the use with sparse LPT data.

Therefore, our first objective here is to prove and quantify the ability of the current technique to detect coherent structures in realistic LPT data. Towards this goal, the current approach is described and tested with synthetic data on the two-dimensional, mixing double-gyre flow, and then subsequently on challenging three-dimensional LPT data behind a bluff body at high Reynolds numbers. Coloured Voronoi cells from the current technique are then compared to CSC-coloured tracks, FTLE-fields, and to diagnostics obtained with the baseline vorticity fields. The outline of this paper is as follows. In section \ref{sec:Methods}, we review the concepts for defining coherence using topological properties of the Voronoi diagrams. This analysis is followed in section \ref{sec:Formulation} through the description of the adopted key normalized parameters for the evaluation of the current method. In section \ref{sec:Benchmarks},  we describe the adopted benchmark flows. In section \ref{sec:Results}, analysis of the technique is performed with changes in the normalized parameter space. Lastly, conditions and limitations for the use of the technique with three-dimensional LPT data are established in section \ref{sec:large_scale_exp}

\section{Methods}
\label{sec:Methods}

The following section is organized as follows. In section \ref{sec:Formulation}, the mathematical description of the proposed method and adopted notation are presented. Subsequently in section \ref{sec:Parameters}, we explain the normalized parameter space for the analysis of the current technique.

\subsection{Mathematical description}
\label{sec:Formulation}

\begin{figure}
  \centering
  \def\svgwidth{\columnwidth}
  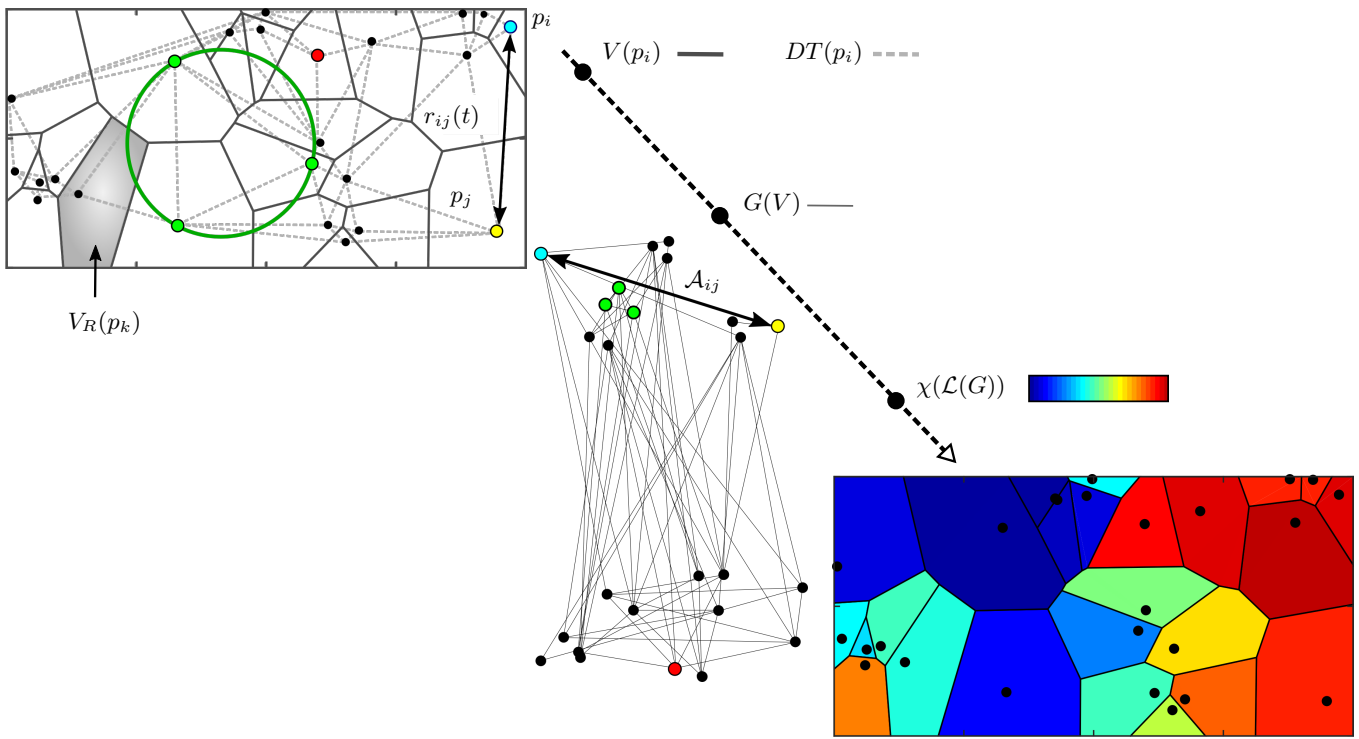
  \caption{Schematics of coherent-structure detection based on Voronoi tessellation. Voronoi and Delaunay diagrams of the particle distribution $p_i$, $V(p_i)$, $DT(p_i)$, are represented in the graph space, $G(V)$, such that lengths of edges of $DT(p_i)$, $r_{ij}$, are converted into distances in the higher-dimensional eigenspace $G(V)$: $r_{ij} \rightarrow \mathcal{A}_{ij} \rightarrow \chi(\mathcal{L}(G))$. A Voronoi cell, $V_R(p_i)$, is highlighted in blue in the top-left panel. The top-left panel also highlights in green the triangulation of three particles intersecting a circumcircle with no particles within its interior.}
\label{fig:Diagram}
\end{figure}

This section presents the mathematical description of the proposed technique. Key steps of coherent-structure detection within the technique's framework are also schematically represented in Fig. \ref{fig:Diagram}. Consider the instantaneous seeding of $N$ particles, $p_i$ ($i \in I_N =\{1,2,\cdots,N\}$), in $m$-dimensional Euclidean space, $m \in I = \{2,3\}$, with location vectors $\mathbf{x}_i$. The Voronoi diagram of $p_i$, $V(p_i)$, is such that for every point $\mathbf{x}$ in space, $\mathbf{x} \in \mathbb{R}^m$, a convex polygon, referred to as a Voronoi cell, $V_R(p_i)$, can be defined as a region in which $\mathbf{x}$ is a closer member to $p_i$ than to any other particle $p_j$ ($p_i \neq p_j$ if $i \neq j$) with location vector $\mathbf{x}_j$:

\begin{equation}
	V_R(p_i) = \left\{ \mathbf{x} \text{  s.t.} \parallel \mathbf{x} - \mathbf{x}_i  \parallel \leq \parallel \mathbf{x} - \mathbf{x}_j  \parallel \text{ for } j \neq i, j \in I_N \right\}.
	\label{Eq:Voronoi_cell}
\end{equation}

In Eq. \ref{Eq:Voronoi_cell}, $\mathbf{x}$ is any point within the fluid domain, and $V_R(p_i)$ encloses each particle into an individual Voronoi cell (see highlighted polygon in Fig. \ref{fig:Diagram}). The boundaries of such a polygon are referred to as Voronoi edges, $r_{ij}$ (see thick black lines in top-left panel in Fig. \ref{fig:Diagram}). The collection of all Voronoi cells at the time instant $t$, $V(p_i)$, given by

\begin{equation}
	V(p_i) = \{ V_R(p_1), \cdots, V_R(p_N) \},
	\label{Eq:Voronoi_diagram}
\end{equation}

\noindent
is the Voronoi diagram of the seeding distribution, $p_i$ \citep{ferrero2011voronoi}. Based on $p_i$, a network called a Delaunay triangulation, $DT(p_i)$, see dashed gray lines in top-left panel in Fig. \ref{fig:Diagram}, can be defined, where the vertices of the network correspond to particle positions, $\mathbf{x}_i$, and a triangle (i.e., a set of three edges) between three adjacent nodes is created if no other particles lay within the circumcircle area defined by the three nodes (see green circle in Fig. \ref{fig:Diagram}). It is evident from the definition of $DT(p_i)$ that Voronoi cells only share a common Voronoi edge when their associated particles are connected by a Delaunay edge \citep{chew1989constrained}. It is also axiomatic from the given definitions of $DT(p_i)$ and $V(p_i)$ that edges of the Voronoi diagram are normal to the Delaunay triangle edges.

Hereafter, we define neighbouring tracers, $p_i$ and $p_j$, all pairs of tracer particles that, at the $n$-th timestep, describe the two vertices of the $(i,j)$-th Delaunay edge (see particles highlighted in green in Fig. \ref{fig:Diagram}). The total neighbouring time, $n_{Tij}$, is subsequently defined as the total number of timesteps, $n_{Tij} \in [0,n]$, for which the $i$-th and $j$-th tracers persisted as neighbouring tracers. Based upon such definitions, the problem of coherent-structure detection is naturally reduced to that of finding groups of Voronoi cells that persist for higher numbers of timesteps as neighbouring cells, since the cells of two particles in a Delaunay edge are neighbouring cells. The present technique solves this last problem by adopting spectral graph theory \citep{spielman2007spectral}. Consider a graph $G(V)$ defined such that its $i$-th node represents the $i$-th Voronoi cell, whereas its $(i,j)$-th edge connects the $i$-th cell to its $j$-th neighbouring Voronoi cell. The weight (or length) $\mathcal{A}_{ij}$ of the $(i,j)$-th edge is defined as

\begin{equation}
  \mathcal{A}_{ij} = \left( \frac{1}{2} \right)^{n_{Tij}},
  \label{Eq:Adjacency}
\end{equation}

\noindent
where $\mathcal{A}$ consists of a full matrix referred to as the adjacency matrix of the graph $G(V)$. From the definition in Eq. \ref{Eq:Adjacency}, it is clear that the length of the $(i,j)$-edge is smaller for greater neighbouring times. Hence, long-lasting neighbouring tracers will be closer in the graph representation of the Voronoi diagram, $G(V)$, and are naturally clustered together. It is also evident that such graphs share the same topology of $DT(p_i)$, as shown in Fig. \ref{fig:Diagram}. Nonetheless, while edges of $DT(p_i)$ yield a physical distance between pairs of tracers, edges of $G(V)$ represent what we define as the \textit{kinematic distance} between the same tracer particles. Moreover, since the neighbouring time is a frame-invariant property, coherent structures obtained with the proposed technique are also objective \citep{haller2000finding}. It is also important to highlight the fact that the lifetime of Voronoi edges is volatile, hence contamination of results by track-breaking process is expected.

Central to the spectral clustering technique are also the definitions of the graph's degree matrix, $\mathcal{D}$, given by

\begin{equation}
  D_{ij} = \delta_{ij} \sum_{k=1}^{N} {A_{ik}},
  \label{Eq:Degree}
\end{equation}

\noindent
where $\delta_{ij}$ represents Kronecker's delta function and the graph's Laplacian, $\mathcal{L}$, defined as

\begin{equation}
  \mathcal{L}_{ij} = \mathcal{D}_{ij}^{-1/2}(\mathcal{D}_{ij}-\mathcal{A}_{ij})\mathcal{D}_{ij}^{-1/2}.
  \label{Eq:Laplacian}
\end{equation}

 In Eq. \ref{Eq:Laplacian}, $\mathcal{L}$ is pre-normalized by $\mathcal{D}$. The corresponding eigenvalue problem that maximizes the differences between the Voronoi cell's neighbouring time is

\begin{equation}
  \mathcal{L}_{ij}X_j = \xi X_j,
  \label{Eq:eigen}
\end{equation}

\noindent
such that $\chi_j \equiv X_j(\xi_{\max})$, $j \in I_N=\{1,\cdots,N\}$, is the eigenvector with associated maximum eigenvalue, $\xi_{\max}$ \citep{schlueter2017identification}. Voronoi cells, $V_R(p_j)$, assigned with similar $\chi_j$ values are supposed to indicate spatio-temporal coherence. This coherence criterion is such that Voronoi cells that share a common Voronoi edge for longer flow times present a shorter distance in the higher-dimensional eigenspace (lower eigengaps), hence presenting coherent motion. In the proposed approach, coherent structures in the fluid flow are identified as regions in the flow with Voronoi cells of similar values of $\chi$ \citep{schlueter2017coherent}.

\subsection{Normalized parameters}
\label{sec:Parameters} 

Considering a volume of fluid that scales with the characteristic length of the flow, $D$, and a homogeneous seeding distribution, accurate coherent-structure identification is assumed to be dependent on the number of tracers, $N$, present in the flow. Consider a flow with largest length scales $D$. Defining the characteristic period of the largest scales as $T$, coherence shall be visualized in normalized time, $t/T$. Moreover, the amount of Lagrangian information in the flow relative to $D$ is assumed to be proportional to the number $N$ of tracer particles tracked per characteristic length $D$, as denoted by $C=N/D^3$. Subsequent to the definition of $C$, here we define the mean inter-particle distance as

\begin{equation}
  \Lambda \equiv \left( \frac{3}{4 \pi C}  \right)^{1/3}.
  \label{Eq:Lambda}
\end{equation}

Accuracy of coherent-structure detection with the proposed approach is assessed by comparing results obtained within the technique's framework with results obtained from the baseline Finite-time Lyapunov exponents (FTLE) approach \citep{haller2000lagrangian} and with the state-of-art Coherent-Structure Colouring (CSC) algorithm \citep{schlueter2017coherent}. Both of these approaches will be summarized in the following paragraphs.

The FTLE measures the maximum linearised growth rate between initially adjacent particles. The FTLE approach is founded on the definition of the Cauchy-green tensor, $\mathcal{C}$, given by

\begin{equation}
  \mathcal{C}(\mathbf{x}_0) = \left( \nabla \mathbf{F}_{t_0}^{t} \right)^\top \cdot \left( \nabla \mathbf{F}_{t_0}^{t} \right),
\end{equation}

\noindent
where $\mathbf{F}_{t_0}^{t}$ represents the flow map, which operates over a fluid particle $p_i$ at $\mathbf{x}_{i,0} \equiv \mathbf{x}_i(t_0)$, $i \in I_N=\{1,\cdots,N\}$, and tracks the particle's position from an initial to a final state, $\mathbf{F}_{t_0}^{t} : \mathbf{x}_0 \rightarrow  \mathbf{x}(\mathbf{x}_0)$. Identification of coherent structures with the FTLE technique relies upon the definition of regions of highest eigenvalue $\lambda_{\max}$:

\begin{equation}
  FTLE_{t_0}^{t} = \frac{1}{|t-t_0|} \ln \left( \sqrt{\lambda_{\max}(\mathcal{C})} \right).
  \label{Eq:FTLE}
\end{equation}

In the FTLE framework, coherent structures are identified as barriers to fluid motion. Extrema values of $FTLE_{t_0}^{t}$ represent the locally more influential material surfaces. If integrated forward in time, extreme positive values represent the locally most repelling material surfaces, whereas opposite in signal extrema represent the most attracting ridges. Both material surfaces exchange definitions if the FTLE is integrated backward in time.

Moreover, $\chi$-coloured Voronoi cells will also be compared to the CSC-coloured tracks from the CSC approach. Similar to the currently-proposed technique, in the CSC framework the dissimilarities between a pair of tracer trajectories is represented by the adjacency matrix of the associated graph-clustering problem, $\mathcal{A}$, defined as

$$ \mathcal{A}_{ij} = \frac{1}{ \overline{r}_{ij} |t-t_0| }
\left[ \sum_n  (\overline{r}_{ij}  - r_{ij}(t_n) )^2 \right]^{1/2}, $$

\noindent
where $\overline{r}_{ij}$ represents the mean distance between the $i$-th and $j$-th particles.  The spectral clustering process based on the CSC approach follows from Eq. \ref{Eq:eigen} such that $CSC \equiv X(\xi_{\max})$. Results obtained with both the FTLE and CSC techniques for varying $\Lambda$-values will be used for comparison.

\section{Description of benchmarks}
\label{sec:Benchmarks}

\subsection{Algorithm verification via case study: double-gyre flow}

We specify an unsteady velocity field $u_i$, $i \in I = \{1,2\}$, the double-gyre flow, that has had extensive use as a test-bed for coherent-structure detection \citep{shadden2005definition,senatore2011detection, williams2015identifying, pratt2015reaction,  balasuriya2016barriers, balasuriya2018generalized}:

\begin{equation}
  \left.\begin{aligned}
    u_1 & = -\pi A \sin\left( \pi h(x_1,t) \right)\cos\left( \pi x_2 \right),\\
    u_2 & =  \pi A  \cos\left( \pi h(x_1,t) \right)\cos\left( \pi x_2 \right)\frac{\partial h}{\partial x_1}(x_1,t) ,\\
        \end{aligned}
  \right\}
  \label{Eq:double_gyre}
\end{equation}

\noindent
where

$$ h(x_1,t) := \epsilon \sin(\varOmega t)x_1^2 + \left( 1-2\epsilon\sin(\varOmega t) \right)x_1.$$

\begin{figure}
  \centering
  \def\svgwidth{\columnwidth}
  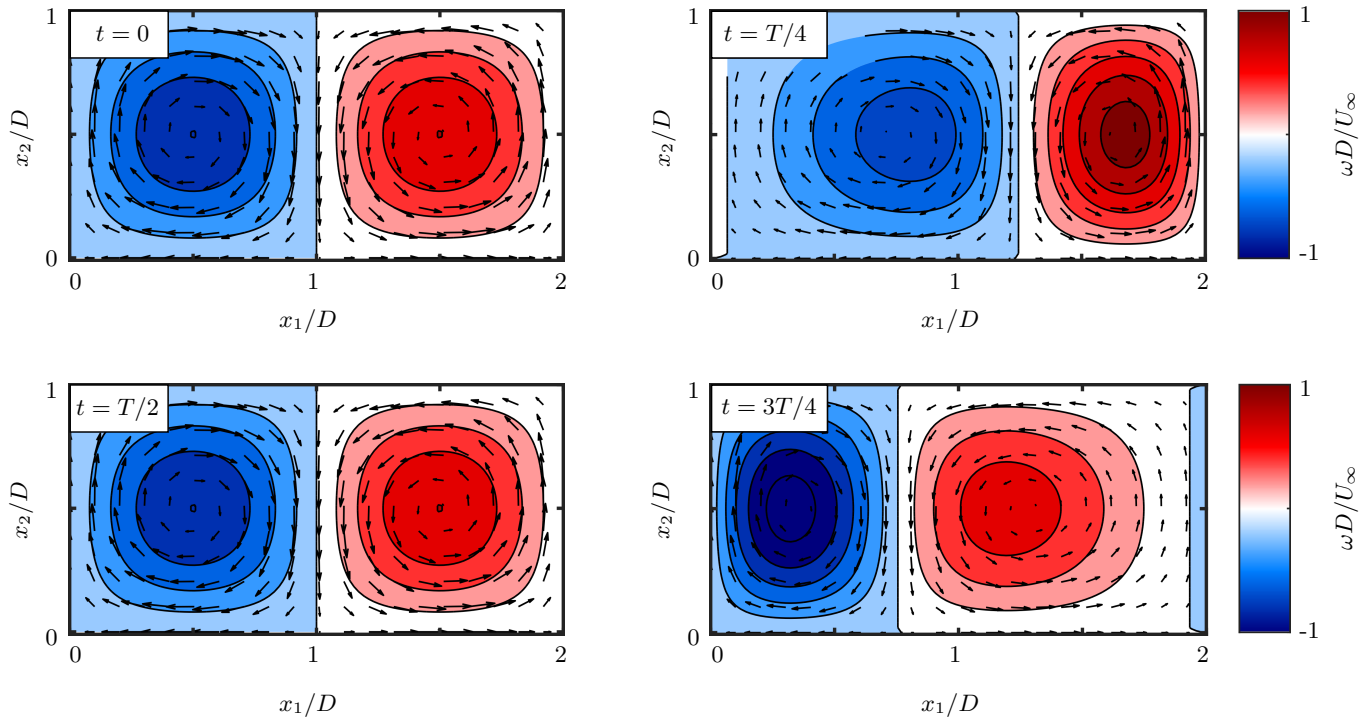  
  \caption{Normalized vorticity fields for the reference double-gyre flow. The plots exhibit contours of normalized vorticity at different timesteps $t$ obtained with a grid of spatial resolution of $\Delta x_i/D = 0.01$.}
  \label{fig:double_gyre}
\end{figure}

For a domain $\Gamma = [0,2] \times [0,1]$, $x_i \in \Gamma$, it is well-known that the topology of the flow comprises hyperbolic trajectories near $x_i = (1,0)$ and $(1,1)$, which move position for $\epsilon>0$ \citep{williams2015identifying}. In the present study, $A=1$, $\epsilon=0.25$ and $\varOmega  = 2\pi/T = 2\pi/10$. The characteristic length $D$ and velocity $U_\infty$ are such that $U_\infty/D=T$. Figure \ref{fig:double_gyre} shows the normalized vorticity field at different characteristic times $T$ for reference. The reference flow is simulated using a pseudo-spectral code with spatial resolution of $\Delta x_i/D = 0.01$, and was evolved with a constant timestep of $\Delta t/T=0.01$.

\subsection{Case study: High-Reynolds-number bluff-body flow}

\begin{figure}
  \centering
  \def\svgwidth{\columnwidth}
  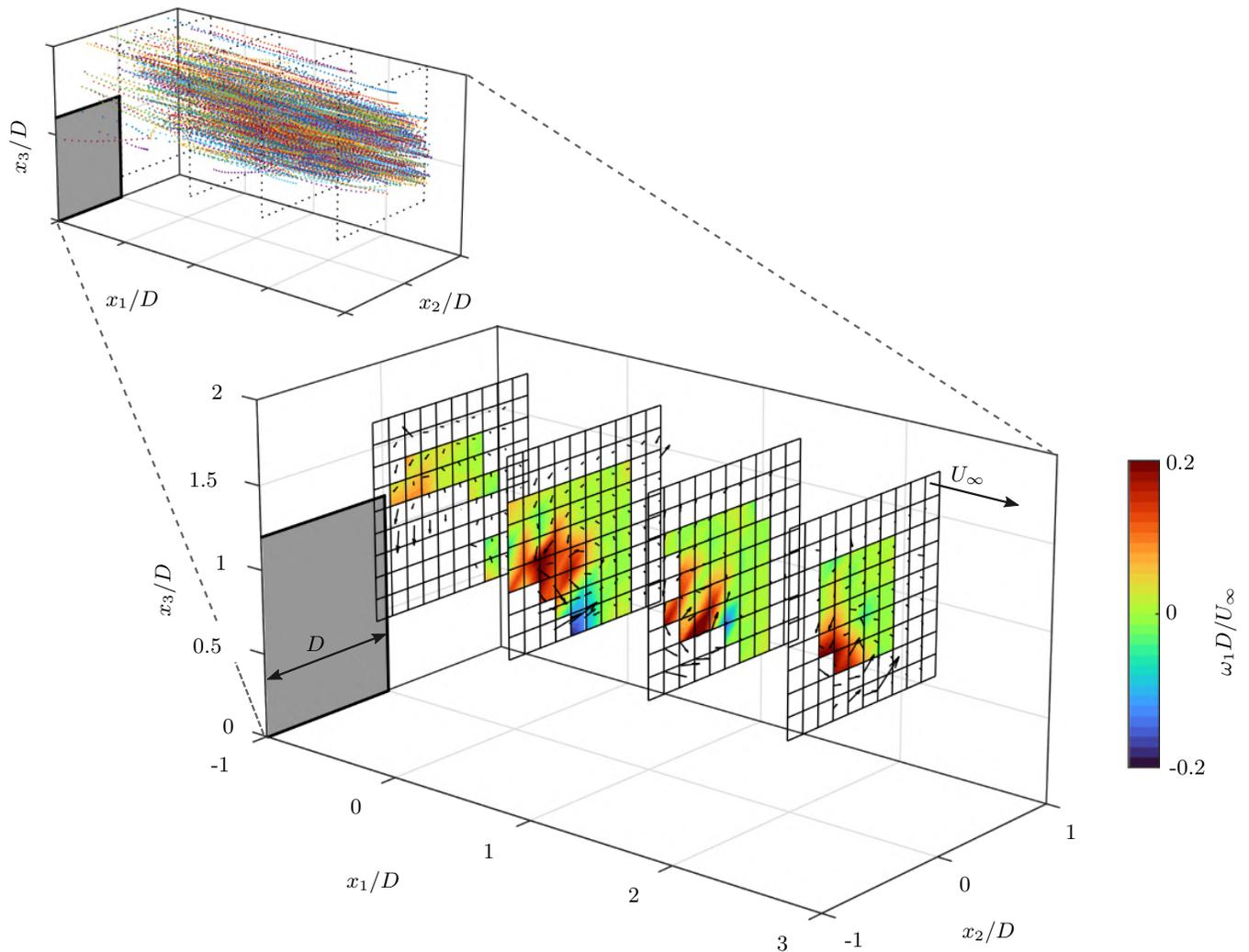  
  \caption{Raw tracks shown in top left. Reference normalized cross-flow vorticity field obtained from the point-in-cell interpolation of the tracer trajectories is shown in the bottom right. Raw reference fields (without smoothing or interpolation) are plotted in specific planes normal to the streamwise direction at $x_1/D = [-0.5,0.5, 1.5, 2.2]$. Results obtained have a constant bin size of $\Delta x_i/D=0.05$. The direction of the reference far-field velocity, $U_\infty$, and the characteristic length, $D$, are also illustrated.}
  \label{fig:large_scale}
\end{figure}

The second benchmark case tested in this study consists of LPT data behind a bluff body collected in the $24\text{ m} \times 9.1\text{ m} \times 9.1\text{ m}$ test section in the large low-speed wind tunnel at the National Research Council in Ottawa, Canada \citep{hou2021novel}. An overview of the raw-tracks data and resultant, interpolated vorticity field, are provided in Fig. \ref{fig:large_scale}. Tracks behind the bluff body are measured using a single-camera setup (Photron mini-WX 100, AF Micro-Nikkor $60\text{ mm}$ $f/2.8$, with the aperture set to 11) in a measurement volume of $4D \times 1.5D \times 1.5D$, where $D$ is the characteristic length scale of the flow ($D \approx 1\text{ m}$). Large air-filled soap bubbles generated with two commercial bubble generators (Antari B200) are adopted as tracer particles, yielding bubbles production rates of $\sim 80 \text{  bubbles/s}$. Soap bubbles with an average diameter of $\sim17.5\text{ mm}$ are illuminated by an array of four pulsed high-power LEDs (LED-Flashlight 300, LaVision GmbH). Characteristic wake-flow behind the bluff body at free-stream velocity of $U_\infty \approx 8\text{ m/s}$, yielding $Re \sim 6 \times 10^5$ based on the model's height, is assessed at a constant sampling frequency of $f_s = 150\text{ Hz}$. More details concerning the measurement process and general flow field are presented in \citet{hou2021novel}.

A total of 5718 bubble tracks are extracted from this large-scale measurement, whereas in the present study we filter out from the raw track data trajectories of temporal length $L_{s,min} \leq 9$ timesteps. The filtering process yields 1940 tracks of temporal length $L_{s,min} \geq 10$ timesteps. Figure \ref{fig:large_scale} represents raw pathlines and a reference-normalized vorticity field obtained from point-in-shell interpolation of the 1940 filtered tracer trajectories. We also highlight the fact that no smoothing or interpolation schemes are adopted to the reference vorticity field shown in Figure \ref{fig:large_scale}.

The analysis of the reference vorticity field indicates that tracks undergo a strong swirling motion due to the C-pillar vortex formed in the wake of the bluff body in yaw. Coherence detection from this LPT data is challenging due to a lack of tracks in the lower-left corner of the measurement volume and in the vortex-core region. Low tracer concentration in the vortex core region is associated with bursting of bubbles due to large pressure and shear-stress gradients near the centerline of the coherent structure. Strong advection outwards from the vortex core also yields inhomogeneity in the tracer distribution. The sparsity and non-homogeneous tracer distribution characteristics of the current data set allow for the assessment of a single C-pillar-like vortex structure in the track data. However, the coherent structure is hardly visible in the vorticity field due to the poor extraction of spatial-gradients achieved with this sparse data set, defined as $\Lambda \geq 0.13 $.

\section{Results and discussion}
\label{sec:Results}

\subsection{Double-gyre flow}

In the following section, the performance of the proposed approach is first assessed on simple, two-dimensional data from the double-gyre flow. Voronoi cells coloured with $\chi$-values are compared to the FTLE- and CSC-coloured fields in a side-by-side fashion. CSC- and FTLE-fields are re-scaled in all cases to the intervals $[-1,1]$ and $[0,1]$, respectively, to allow for direct comparison of results. We extract coherent structures using the proposed technique by simply re-scaling $\chi$ to the interval of $[-1,1]$.

The first analysis consists of comparing results obtained with the three approaches for decreasing mean inter-particle distance values, whereas the normalized flow time is fixed at $t/T=2$. Subsequently, the same analysis is repeated for a comparison of the current technique to the baseline vorticity field. Here, the objective is to perform a systematic comparison of the different techniques, using them on a test-bed flow in which a ground truth can be reasonably established. Nevertheless, due to the absence of a standardized method of quantifying the quality of coherent-structure detection techniques that are fundamentally different in nature, diagnostic tools in this study are compared purely through visual inspection.

In the present analysis, the number of tracers, $N$, ranges considerably ($N=36$, $64$, $100$, $196$, $484$ and $1225$), corresponding to mean inter-particle distances of $\Lambda \approx 0.24$, $0.20$, $0.17$, $0.13$, $0.10$ and $0.07$. In a recent publication, it was concluded that particle concentrations of $N/D^3 \leq \mathcal{O}(10^2)$ ($\Lambda \geq 0.13$) are insufficient for coherent-structure detection in boundary-layer or free-jet flows based on LPT data \citep{schneiders2016dense}. Hence, here we consider particle distributions of $\Lambda \geq 0.13$ as very sparse data sets. Tracers are homogeneously-seeded using the following procedure: in every test case, the domain $\Gamma_i$, $i \in I = \{1,2\}$, is split into bins of size $\Delta x_1 \times \Delta x_2 = \Gamma_1/N^{1/2} \times \Gamma_2/N^{1/2}$, followed by the seeding of a single particle into a random position inside one of the $N$ uniformly-distributed bins. After the homogeneous seeding of synthetic tracers, tracer trajectories are integrated using a second-order central scheme with timesteps of $\Delta t/T=0.1$ in $t/T \in [0,10]$. Test cases are summarized in Table \ref{tab:summary_dg}.

\begin{table}
\begin{ruledtabular}
    \caption{Summary of test cases for the double-gyre flow, where $N$ is the number of tracer particles and $\Lambda$ is the mean inter-particle distance.}
    \begin{tabular}{ccccccc}
    Case             & a    & b    & c    & d    & e    & f    \\
    \hline
    $N$              & 36   & 64   & 100  & 196  & 484  & 1225  \\
    $\Lambda\approx$ & 0.24 & 0.20 & 0.17 & 0.13 & 0.10 & 0.07 \\
    \end{tabular}
  \label{tab:summary_dg}
\end{ruledtabular}
\end{table}

\begin{figure}
  \centering
  \def\svgwidth{\columnwidth}
  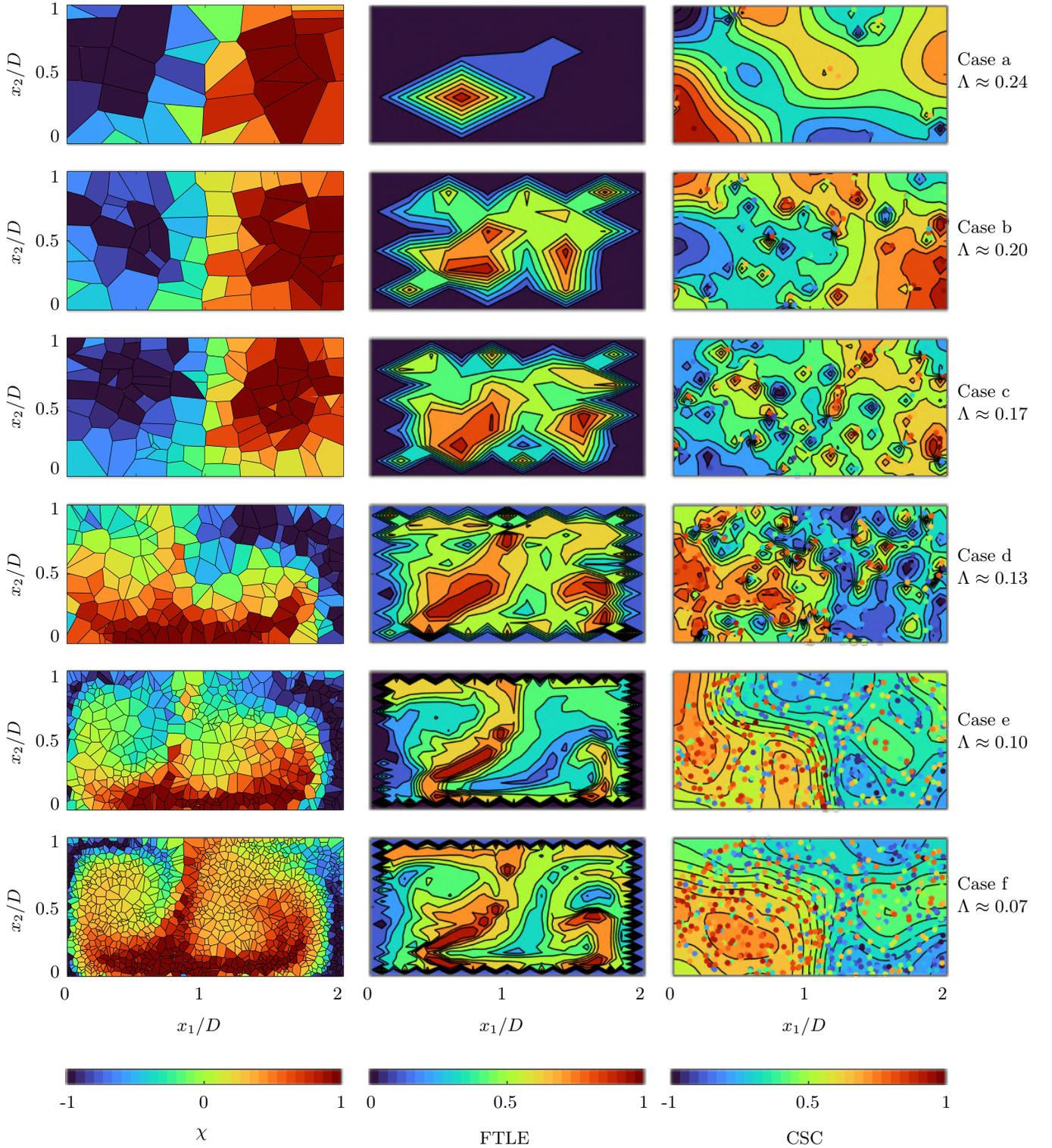  
  \caption{Selected cases for double-gyre flow at $t=2T$ with mean inter-particle distance values of $\Lambda\approx0.24, 0.20, 0.17,  0.13, 0.10, 0.07$ (top-to-bottom). Analyses are visualized by $\chi$-coloured Voronoi cells (left), FTLE-fields (center) and CSC-coloured fields overlaid by CSC-coloured tracks (right). Cases (a) to (f) are listed in Table \ref{tab:summary_dg}.}
  \label{fig:double_gyre_results}
\end{figure}

Figure \ref{fig:double_gyre_results} presents the double-gyre flow for, from left-to-right, $\chi$-coloured Voronoi cells, FTLE- and CSC-fields. FTLE fields were computed by a Matlab software package developed by \citet{peng2009transport}. The algorithm can be summarized as follows: a third-order interpolation scheme is adopted to interpolate the flow map function, $\mathbf{F}_{t_0}^{t}$, onto a grid. Subsequently, spatial gradients are obtained through central differencing with neighbouring grid points. Lastly, the FTLE field is computed at every grid point based of Eq. \ref{Eq:FTLE}. CSC-fields are obtained through the mapping of the instantaneous seeding distribution onto two-dimensional Cartesian grids. The resultant field is then subjected to a robust spline smoothing algorithm developed by \citet{garcia2010robust}, which adopts a type-2 discrete cosine transform to assign missing data to values that are estimated using the entire dataset.  Moreover, we use this blue-green-yellow-red colour-scheme consistently for all diagnostic fields for ease of comparison. From the results shown in Fig. \ref{fig:double_gyre_results}, is it observed that without requiring specification of the number of gyres, the proposed technique was able to reveal physically-interpretable coherent regions for the sparsest test cases, at $\Lambda \approx 0.24$ and $\Lambda \approx 0.20$ (see cases (a) and (b), respectively, in Fig. \ref{fig:double_gyre_results}). Despite such low particle concentration, the two gyres can be identified in the flow as two regions in the Voronoi diagram coloured in red and blue, corresponding to the two extrema of $\chi$. In this case, the largest dissimilarity in the flow in terms of neighbouring time is between trajectories that started near the left and right quadrants of the double-gyre flow. At this same particle concentration, both the CSC and FTLE techniques were unable to provide physically-interpretable results.

Further decreasing the inter-particle distance to $\Lambda\approx0.17$, for case (c), the boundary between the two distinguished regions in red and blue in the Voronoi diagram becomes well-defined, whereas no discernible pattern has emerged from the FTLE or CSC approaches. Two regions start to form in the CSC-fields only further decreasing the mean inter-particle distance to $\Lambda \approx 0.10$, for case (e), two regions start to form in the CSC-fields, which are evident from the two clusters of tracers somewhat reproducing the expected topologies of the two gyres. The fact that material coherence can be observed in CSC-fields with the particle concentration of case (e) agrees with observations of \citet{husic2019simultaneous}, who concluded that the CSC is expected to accurately reconstruct the two gyres of the double-gyre flow with a minimum number of tracks of $N\sim300$, or $\mathcal{O}(\Lambda) \leq 0.1$, representing a considerable progress towards sparser data sets when compared to classical gradient-based approaches.

By decreasing Lambda (cases (d), (e), and (f) of Fig. \ref{fig:double_gyre_results}) we observe the appearance of a different colour distribution for $\chi$. In fact, the two gyres are mainly characterized by intermediate $\chi$ values ($\chi \approx 0$) while the strongest (blue and red) $\chi$ values surround the two gyres. These first results reveal an important aspect of the tested technique: if sufficient particle concentration is provided, i.e., $\mathcal{O}(\Lambda/D) \ll 1$, dissimilar colours will inevitably be attributed to regions that act as barriers to fluid motion, since more particles tend to cluster in such regions, thus showing high neighbouring times. In cases of sparse particle concentrations (such as in cases (a) and (b)), $\mathcal{O}(\Lambda/D) \sim 1$, the dissimilarities will be simply attributed to tracks that were advected by the strongest coherent structures. Equivalently, such a result reveals that Voronoi cells coloured in dark blue rarely interact with tracks coloured in dark red. On the other hand, green regions could be expected to represent regions of more intense mixing. In the context of transport of passive scalars, for example, one should expect little to no mixing of the transported quantity between regions of the flow coloured with $\chi_{\max}$ and $\chi_{\min}$, which is a useful diagnostic for many natural and engineering flows. We also note that there is an analogy with FTLE field for low $\Lambda$ values, such that high $\chi$ values share a similar topology to the unstable manifold (high FTLE values) of the FTLE technique (see cases (e) and (f) in Fig. \ref{fig:double_gyre_results}, for example).

We repeat the comparative analysis using an Eulerian vortex criterion. Due to the simple topology of the flow, the vorticity field, $\bm{\omega} = \nabla \times \mathbf{u}$, is sufficient for this study. The $\omega$-fields are obtained as follows. The domain $\Gamma_i$ is divided into bins of size $\Delta x_i/D = 0.05$. Velocity components of the $(i,j)$-th bin are computed using the average velocity of the tracers within its domain, in a point-in-cell fashion \citep{garth2010fast}, whereas empty cells are assigned not-a-number (NaN). After mapping the reconstructed Lagrangian velocity to Eulerian space, the curl of the interpolated velocity field is computed such that partial derivatives are calculated using a second-order central difference scheme for interior data points, whereas first-order, single-sided (forward) difference schemes are adopted for points along the edges \citep{moukalled2016finite}. $\omega$-fields are also subjected to a partial reconstruction process using a linear interpolation algorithm \citep{henn2013comparison} that replaces NaNs in interior points by the arithmetic mean of the respective field variable in the adjacent cells.

\begin{figure}
  \centering
  \def\svgwidth{\columnwidth}
  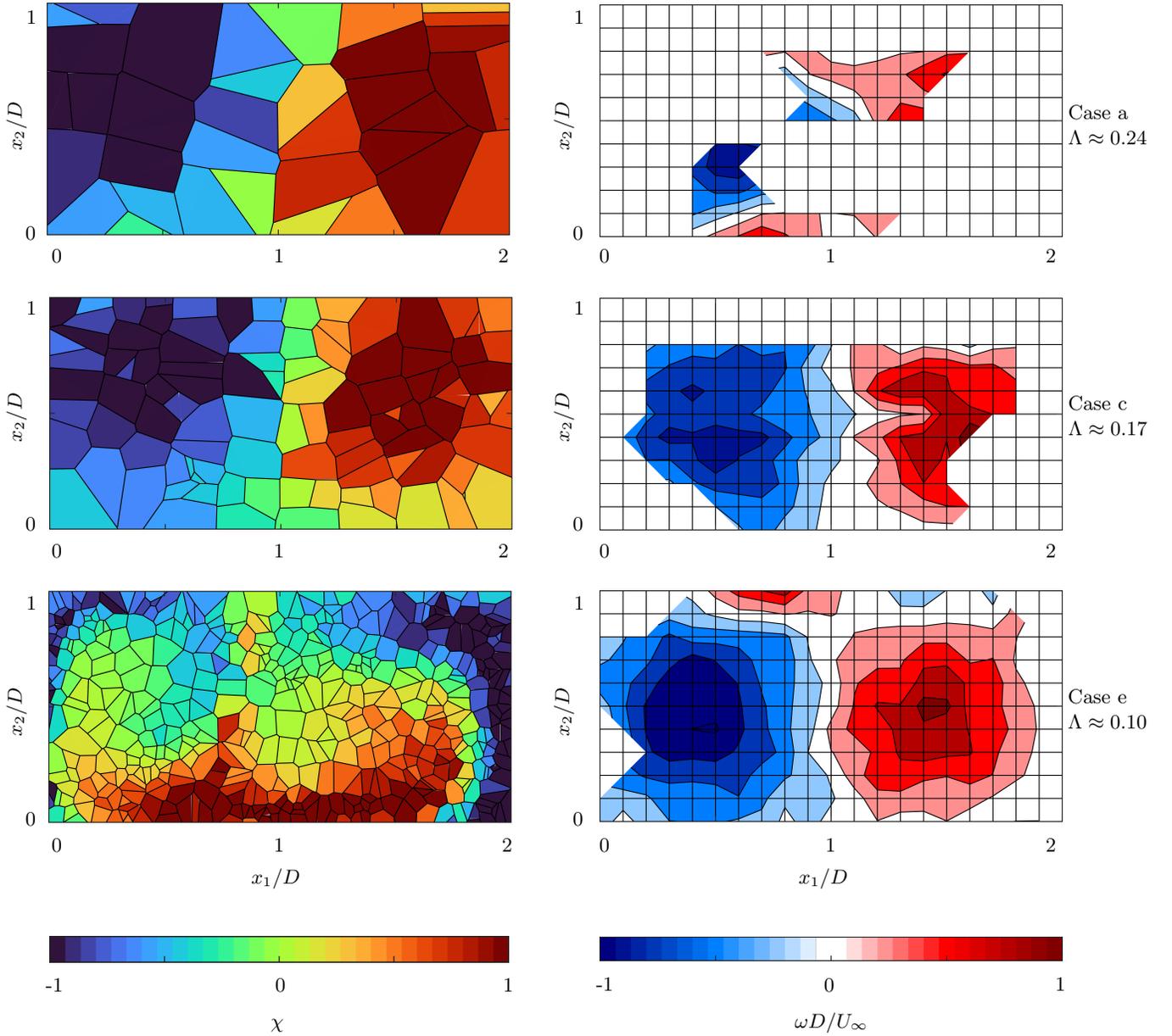  
  \caption{Selected cases of double-gyre flow at $t=2T$ with mean inter-particle distance values of $\Lambda \approx 0.24, 0.17, 0.10$ (top to bottom) visualized by $\chi$-coloured Voronoi cells (left) and $\bm{\omega}$-field (right).}
  \label{fig:double_gyre_results_eulerian}
\end{figure}

Using the Eulerian approach for case (a), shown in the top-right panel of Fig. \ref{fig:double_gyre_results_eulerian}, the very sparse Lagrangian data (36 particles) does not allow for the velocity field to be reconstructed, as expected, whereas the two gyres can be identified as two clusters of Voronoi cells when using the current approach. For $\Lambda \approx 0.17$, corresponding to case (c), a noisy vorticity-field can be reconstructed from the LPT data. Concomitantly, the two gyres are physically distinguishable in the Voronoi diagram. Finally, for  $\Lambda \approx 0.07$, the strongest ridge is observed in the Voronoi diagram in the bottom panels of Fig. \ref{fig:double_gyre_results_eulerian}. At this particle concentration level, the two counter-rotating gyres are also easily distinguishable in the vorticity field.

\subsection{High-Reynolds-number bluff-body data}
\label{sec:large_scale_exp}

The second benchmark conducted in this study consists of three-dimensional LPT data collected behind a bluff body at a free-stream velocity of $U_\infty \approx 8\text{ m/s}$ ($Re \sim 6 \times 10^5$). In the current study, the total number of tracks in each test case is selected by imposing thresholded temporal track lengths of $L_{s,min}$, such that each analysed case consists of tracks of $L_{s} \geq L_{s,min}$ only. The analysis is conducted with $N=1940$, 289 and 70 particles, equivalent to inter-particle distances of $\Lambda \approx 0.065$, 0.12 and 0.20, respectively. We concluded from a preliminary analysis that both the CSC and FTLE techniques fail to provide physically-distinguishable flow features with such challenging LPT data and hence in the current benchmark, results obtained with the proposed technique are compared only to baseline vorticity fields.

In contrast with the two-dimensional Voronoi tessellation, three-dimensional Voronoi diagrams present complex topologies and make the visualization of coherent structures in LPT data quite challenging. We thus opted to omit the Voronoi diagram from the current section as well. Instead, the three-dimensional track data is colour-coded directly with the corresponding $\chi_i$ values. Subsequently, a point-in-cell interpolation is adopted to map the $\chi$-coloured tracks onto three-dimensional $\chi$-fields. The results of the proposed approach are then compared to normalized cross-flow vorticity fields, $\omega_1 D/U_\infty$. Both $\chi$- and $\omega$-fields are analyzed at three planes parallel to the cross-flow direction, at $x_1/D=[-0.5, 1.25, 2.5]$. A constant bin size of $\Delta x_i/D=0.05$ is adopted for the interpolation of the velocity- and $\chi$-fields, which are subsequently subjected to a partial reconstruction process using the robust smoothing algorithm developed by \cite{garcia2010robust}.

\begin{figure}
  \centering
  \def\svgwidth{\columnwidth}
  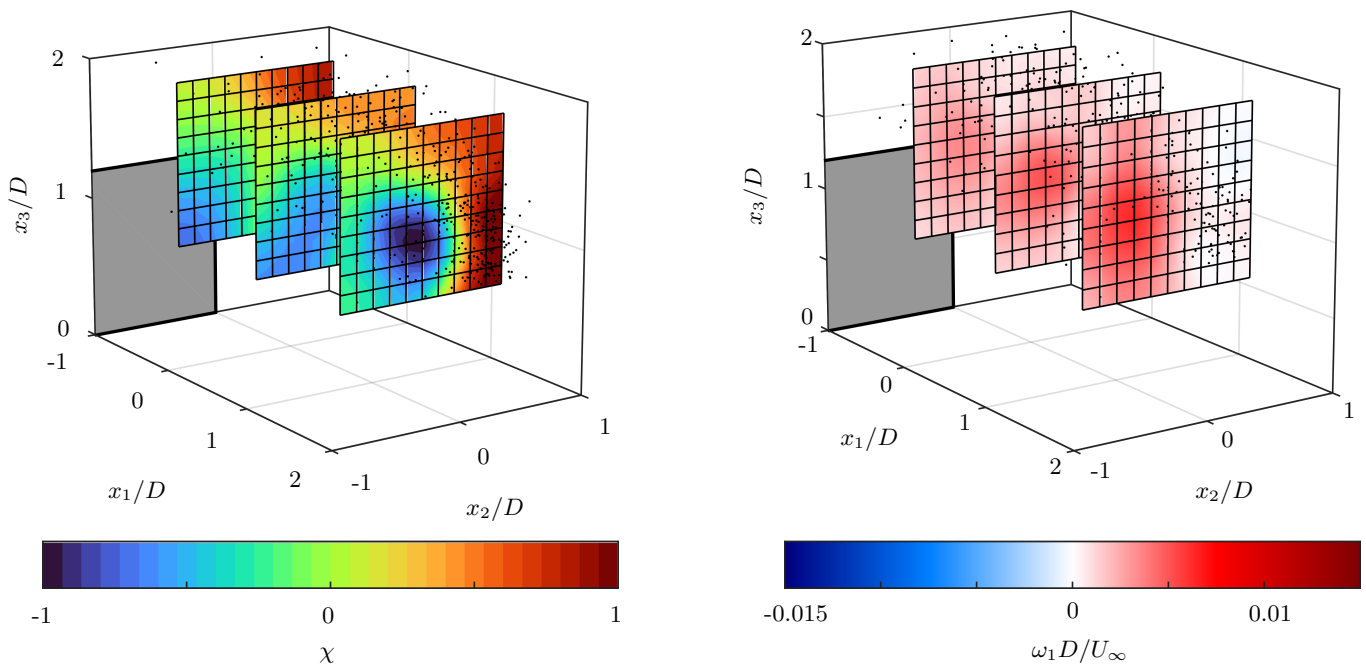  
  \caption{Left and right panels show $\chi$- and vorticity fields, respectively, obtained from the point-in-cell interpolation of the LPT data onto a grid. Results obtained with $\Lambda\approx 0.065$ at $x_1/D=[-0.5, 1.25, 2.5]$. Panel on left also shows $\chi$-coloured tracks.}
  \label{fig:large_scale_results_dense}
\end{figure}

Results for the challenging three-dimensional bluff-body data with $\Lambda\approx0.065$ are shown in Fig. \ref{fig:large_scale_results_dense}, in which the left panel shows $\chi$-fields overlaid by $\chi$-coloured tracks of the current technique, and the right panel shows the vorticity field resulting from the mapping of LPT data onto a grid. As guaranteed by the analysis conducted with the double-gyre flow, the vortex boundaries of the C-pillar vortex are well-defined in the $\chi$-coloured tracks with the selected inter-particle distance value. In contrast, the vortex core, which is located near the center of these planes, can only be partially-visualized by the vorticity field.

\begin{figure}
  \centering
  \def\svgwidth{\columnwidth}
  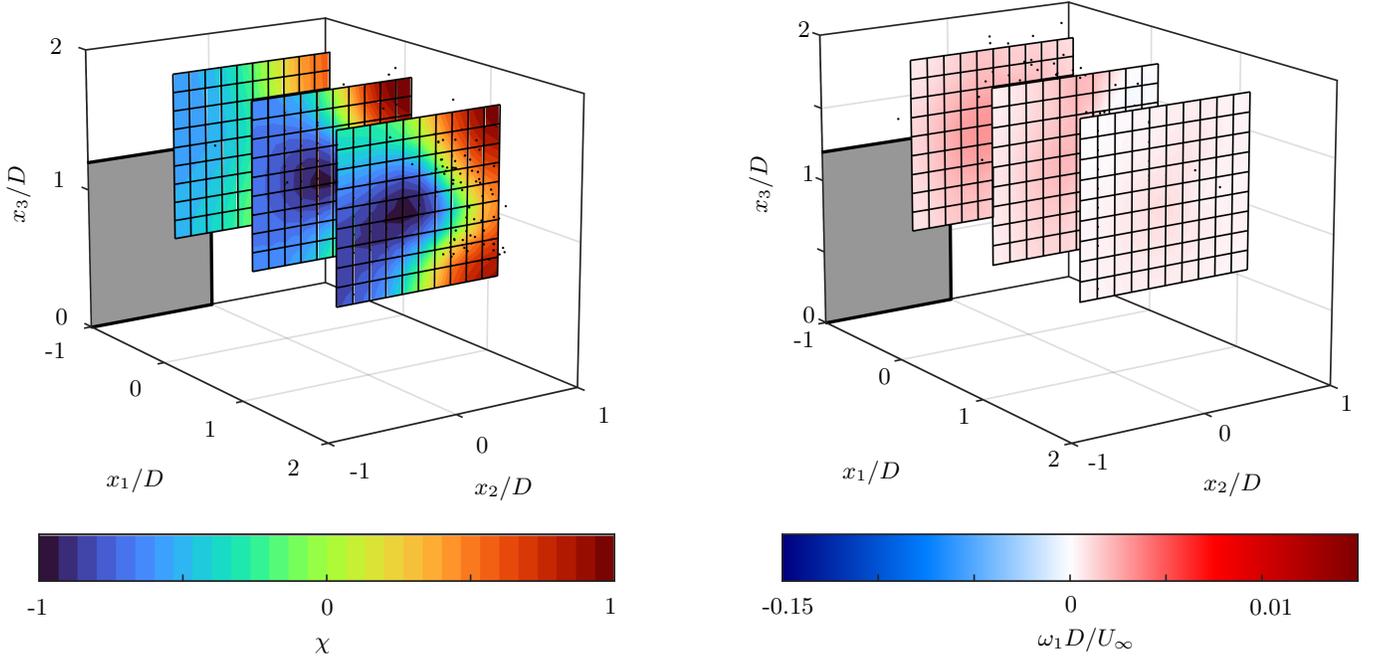  
  \caption{Left and right panels show $\chi$- and vorticity fields, respectively, obtained from the point-in-cell interpolation of the LPT data onto a grid. Results obtained with $\Lambda\approx 0.12$ at $x_1/D=[-0.5, 1.25, 2.5]$. Panel on left also shows $\chi$-coloured tracks.}
  \label{fig:large_scale_results_medium}
\end{figure}

Results for $\Lambda\approx0.12$ are shown in Fig. \ref{fig:large_scale_results_medium}. With a mean inter-particle distance of about one-tenth of the characteristic length scale of the flow, $\Lambda/D \sim \mathcal{O}(10^{-1})$, the Eulerian approach (right panel) lacks a nested sequence of smooth closed contours that allows for coherent-structure detection. Nevertheless, an intuitive visual inspection of the $\chi$-field in the left panel still picks up the convex, C-pillar vortex core at the center of the $\chi$-coloured track data.

\begin{figure}
  \centering
  \def\svgwidth{\columnwidth}
  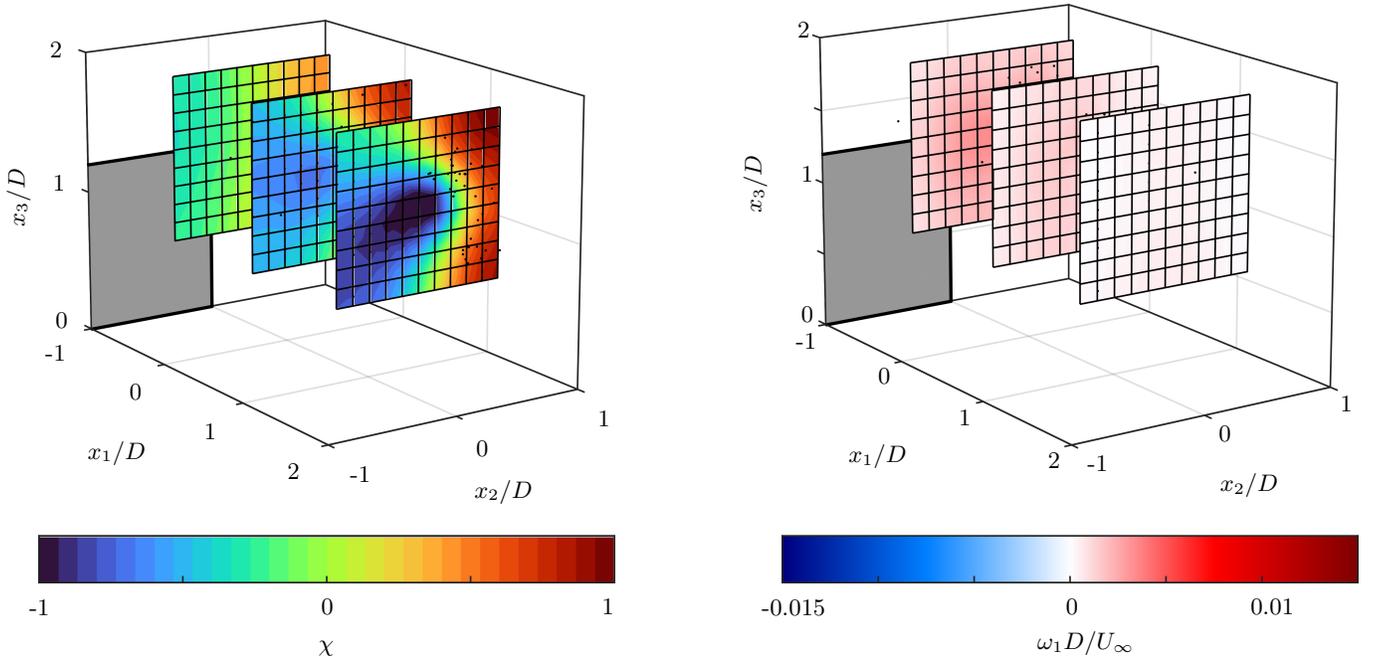  
  \caption{Left and right panels show $\chi$- and vorticity fields, respectively, obtained from the point-in-cell interpolation of the LPT data onto a grid. Results obtained with $\Lambda\approx 0.20$ at $x_1/D=[-0.5, 1.25, 2.5]$. Panel on left also shows $\chi$-coloured tracks.}
  \label{fig:large_scale_results_coarse}
\end{figure}

Lastly, results for $\Lambda\approx0.2$ are shown in Fig. \ref{fig:large_scale_results_coarse}. Interpolation of the velocity field with a large $\Lambda$ value is naturally unreliable, indicated by the poorly-reconstructed vorticity field in the right panel of the figure. On the other hand, as shown in the left panel, boundaries of a convex structure representing the C-pillar vortex can still be partially visualized by the $\chi$-fields, indicating that coherent structures can be identified with the current technique despite the severe sparsity of the investigated experimental data set. Notwithstanding, due to the evident sparsity and non-homogeneous tracer particle distribution of the current data set, it is challenging at this stage to go much beyond a qualitative assessment of the results. Nonetheless, the current technique, in view of the results presented in this study, proves to be a strong candidate for the detection of coherent structures in realistic, very sparse LPT data.

It is important to highlight the fact that the analysed flow is a very challenging benchmark due to data gaps near the core of the C-pillar vortex. Data gaps in the interior of a vortex structure are a consequence of the intense drifting away of tracers from the vortex core due to centrifugal forces \citep{marshall2005particle}. In the analysed data set, strong shear is also responsible for the bursting of tracers (helium-filled soap bubbles) near the center of the C-pillar vortex \citep{faleiros2019generation}. The combination of both effects causes radial-dispersion of tracers and data-gaps along the vortex core, omitting or strongly dampening the coherent structure in diagnostic fields of gradient-based approaches, despite the average particle concentration of the data set (see vorticity fields in Fig. \ref{fig:large_scale_results_dense}, for example). Unlike the neighbouring times of particles advected by the mean flow, which are typically high, the complex advection processes of tracers in regions of high shear yields low neighbouring times. Differently from classical approaches, such contrast between the behaviour of particles inside and outside of vortical regions allows for the visualization of coherent structures using the tested technique even if the particle concentration in the vortex-core region is considerably lower (see results in Fig. \ref{fig:large_scale_results_dense}). Going forward, in the future development of the technique analysed in this study, concrete connections between specific values of $\chi$, and the repelling or attracting behaviour of the coherent structures, will have to be explored.

\section{Conclusions}
\label{sec:Conclusions}

A Voronoi-tessellation-based approach for detection of coherent structures in sparsely-seeded flows is tested. Neighbouring time of tracer trajectories, defined as the total flow time two Voronoi cells remain connected by a Voronoi edge, was adopted as a metric for coherence. The novel approach was tested with LPT data from a simple double-gyre flow and then with challenging LPT data behind a bluff body. The method was compared to CSC-coloured tracks of the CSC technique, FTLE ridges and to the baseline vorticity field. In this study, capabilities of coherent-structures visualization was evaluated using as a metric the inter-particle distance. It was concluded that, in general, higher accuracy in coherent-structure identification is obtained for decreasing inter-particle distances. For low inter-particle distance values, colour-coded Voronoi cells of the current technique were found to exhibit patterns that are strongly correlated to the attracting surfaces (unstable manifolds) of the FTLE technique. Moreover, the tested technique demonstrated great potential for applications of inter-particle distances, $\Lambda$, on the order of the characteristic length scale $D$ of the flow, i.e., $\Lambda \sim \mathcal{O}(D)$, whereas other techniques present limited value.

\begin{acknowledgments}
  DER acknowledges funding from the Early Researcher Award (Ontario). The authors would also like to thank Dr. G. Iacobello for the many insightful discussions and feedback on the manuscript.
\end{acknowledgments}

\nocite{*}
\bibliography{bibliography}

\begin{thebibliography}{43}%
\makeatletter
\providecommand \@ifxundefined [1]{%
 \@ifx{#1\undefined}
}%
\providecommand \@ifnum [1]{%
 \ifnum #1\expandafter \@firstoftwo
 \else \expandafter \@secondoftwo
 \fi
}%
\providecommand \@ifx [1]{%
 \ifx #1\expandafter \@firstoftwo
 \else \expandafter \@secondoftwo
 \fi
}%
\providecommand \natexlab [1]{#1}%
\providecommand \enquote  [1]{``#1''}%
\providecommand \bibnamefont  [1]{#1}%
\providecommand \bibfnamefont [1]{#1}%
\providecommand \citenamefont [1]{#1}%
\providecommand \href@noop [0]{\@secondoftwo}%
\providecommand \href [0]{\begingroup \@sanitize@url \@href}%
\providecommand \@href[1]{\@@startlink{#1}\@@href}%
\providecommand \@@href[1]{\endgroup#1\@@endlink}%
\providecommand \@sanitize@url [0]{\catcode `\\12\catcode `\$12\catcode
  `\&12\catcode `\#12\catcode `\^12\catcode `\_12\catcode `\%12\relax}%
\providecommand \@@startlink[1]{}%
\providecommand \@@endlink[0]{}%
\providecommand \url  [0]{\begingroup\@sanitize@url \@url }%
\providecommand \@url [1]{\endgroup\@href {#1}{\urlprefix }}%
\providecommand \urlprefix  [0]{URL }%
\providecommand \Eprint [0]{\href }%
\providecommand \doibase [0]{https://doi.org/}%
\providecommand \selectlanguage [0]{\@gobble}%
\providecommand \bibinfo  [0]{\@secondoftwo}%
\providecommand \bibfield  [0]{\@secondoftwo}%
\providecommand \translation [1]{[#1]}%
\providecommand \BibitemOpen [0]{}%
\providecommand \bibitemStop [0]{}%
\providecommand \bibitemNoStop [0]{.\EOS\space}%
\providecommand \EOS [0]{\spacefactor3000\relax}%
\providecommand \BibitemShut  [1]{\csname bibitem#1\endcsname}%
\let\auto@bib@innerbib\@empty
\bibitem [{\citenamefont {Peacock}\ and\ \citenamefont
  {Dabiri}(2010)}]{peacock2010introduction}%
  \BibitemOpen
  \bibfield  {author} {\bibinfo {author} {\bibfnamefont {T.}~\bibnamefont
  {Peacock}}\ and\ \bibinfo {author} {\bibfnamefont {J.}~\bibnamefont
  {Dabiri}},\ }\href@noop {} {\bibinfo {title} {Introduction to focus issue:
  Lagrangian coherent structures}} (\bibinfo {year} {2010})\BibitemShut
  {NoStop}%
\bibitem [{\citenamefont {Hadjighasem}\ \emph {et~al.}(2017)\citenamefont
  {Hadjighasem}, \citenamefont {Farazmand}, \citenamefont {Blazevski},
  \citenamefont {Froyland},\ and\ \citenamefont
  {Haller}}]{hadjighasem2017critical}%
  \BibitemOpen
  \bibfield  {author} {\bibinfo {author} {\bibfnamefont {A.}~\bibnamefont
  {Hadjighasem}}, \bibinfo {author} {\bibfnamefont {M.}~\bibnamefont
  {Farazmand}}, \bibinfo {author} {\bibfnamefont {D.}~\bibnamefont
  {Blazevski}}, \bibinfo {author} {\bibfnamefont {G.}~\bibnamefont
  {Froyland}},\ and\ \bibinfo {author} {\bibfnamefont {G.}~\bibnamefont
  {Haller}},\ }\bibfield  {title} {\bibinfo {title} {A critical comparison of
  lagrangian methods for coherent structure detection},\ }\href@noop {}
  {\bibfield  {journal} {\bibinfo  {journal} {Chaos: An Interdisciplinary
  Journal of Nonlinear Science}\ }\textbf {\bibinfo {volume} {27}},\ \bibinfo
  {pages} {053104} (\bibinfo {year} {2017})}\BibitemShut {NoStop}%
\bibitem [{\citenamefont {Haller}\ and\ \citenamefont
  {Yuan}(2000)}]{haller2000lagrangian}%
  \BibitemOpen
  \bibfield  {author} {\bibinfo {author} {\bibfnamefont {G.}~\bibnamefont
  {Haller}}\ and\ \bibinfo {author} {\bibfnamefont {G.}~\bibnamefont {Yuan}},\
  }\bibfield  {title} {\bibinfo {title} {Lagrangian coherent structures and
  mixing in two-dimensional turbulence},\ }\href@noop {} {\bibfield  {journal}
  {\bibinfo  {journal} {Physica D: Nonlinear Phenomena}\ }\textbf {\bibinfo
  {volume} {147}},\ \bibinfo {pages} {352} (\bibinfo {year}
  {2000})}\BibitemShut {NoStop}%
\bibitem [{\citenamefont {Froyland}\ and\ \citenamefont
  {Padberg-Gehle}(2015)}]{froyland2015rough}%
  \BibitemOpen
  \bibfield  {author} {\bibinfo {author} {\bibfnamefont {G.}~\bibnamefont
  {Froyland}}\ and\ \bibinfo {author} {\bibfnamefont {K.}~\bibnamefont
  {Padberg-Gehle}},\ }\bibfield  {title} {\bibinfo {title} {A rough-and-ready
  cluster-based approach for extracting finite-time coherent sets from sparse
  and incomplete trajectory data},\ }\href@noop {} {\bibfield  {journal}
  {\bibinfo  {journal} {Chaos: An Interdisciplinary Journal of Nonlinear
  Science}\ }\textbf {\bibinfo {volume} {25}},\ \bibinfo {pages} {087406}
  (\bibinfo {year} {2015})}\BibitemShut {NoStop}%
\bibitem [{\citenamefont {Banisch}\ and\ \citenamefont
  {Koltai}(2017)}]{banisch2017understanding}%
  \BibitemOpen
  \bibfield  {author} {\bibinfo {author} {\bibfnamefont {R.}~\bibnamefont
  {Banisch}}\ and\ \bibinfo {author} {\bibfnamefont {P.}~\bibnamefont
  {Koltai}},\ }\bibfield  {title} {\bibinfo {title} {Understanding the geometry
  of transport: Diffusion maps for lagrangian trajectory data unravel coherent
  sets},\ }\href@noop {} {\bibfield  {journal} {\bibinfo  {journal} {Chaos: An
  Interdisciplinary Journal of Nonlinear Science}\ }\textbf {\bibinfo {volume}
  {27}},\ \bibinfo {pages} {035804} (\bibinfo {year} {2017})}\BibitemShut
  {NoStop}%
\bibitem [{\citenamefont {Padberg-Gehle}\ and\ \citenamefont
  {Schneide}(2017)}]{padberg2017network}%
  \BibitemOpen
  \bibfield  {author} {\bibinfo {author} {\bibfnamefont {K.}~\bibnamefont
  {Padberg-Gehle}}\ and\ \bibinfo {author} {\bibfnamefont {C.}~\bibnamefont
  {Schneide}},\ }\bibfield  {title} {\bibinfo {title} {Network-based study of
  lagrangian transport and mixing},\ }\href@noop {} {\bibfield  {journal}
  {\bibinfo  {journal} {Nonlinear Processes in Geophysics}\ }\textbf {\bibinfo
  {volume} {24}},\ \bibinfo {pages} {661} (\bibinfo {year} {2017})}\BibitemShut
  {NoStop}%
\bibitem [{\citenamefont {Schmale~III}\ and\ \citenamefont
  {Ross}(2015)}]{schmale2015highways}%
  \BibitemOpen
  \bibfield  {author} {\bibinfo {author} {\bibfnamefont {D.~G.}\ \bibnamefont
  {Schmale~III}}\ and\ \bibinfo {author} {\bibfnamefont {S.~D.}\ \bibnamefont
  {Ross}},\ }\bibfield  {title} {\bibinfo {title} {Highways in the sky: Scales
  of atmospheric transport of plant pathogens},\ }\href@noop {} {\bibfield
  {journal} {\bibinfo  {journal} {Annual review of phytopathology}\ }\textbf
  {\bibinfo {volume} {53}} (\bibinfo {year} {2015})}\BibitemShut {NoStop}%
\bibitem [{\citenamefont {Shuckburgh}\ \emph {et~al.}(2009)\citenamefont
  {Shuckburgh}, \citenamefont {Jones}, \citenamefont {Marshall},\ and\
  \citenamefont {Hill}}]{shuckburgh2009robustness}%
  \BibitemOpen
  \bibfield  {author} {\bibinfo {author} {\bibfnamefont {E.}~\bibnamefont
  {Shuckburgh}}, \bibinfo {author} {\bibfnamefont {H.}~\bibnamefont {Jones}},
  \bibinfo {author} {\bibfnamefont {J.}~\bibnamefont {Marshall}},\ and\
  \bibinfo {author} {\bibfnamefont {C.}~\bibnamefont {Hill}},\ }\bibfield
  {title} {\bibinfo {title} {Robustness of an effective diffusivity diagnostic
  in oceanic flows},\ }\href@noop {} {\bibfield  {journal} {\bibinfo  {journal}
  {Journal of physical oceanography}\ }\textbf {\bibinfo {volume} {39}},\
  \bibinfo {pages} {1993} (\bibinfo {year} {2009})}\BibitemShut {NoStop}%
\bibitem [{\citenamefont {Davis}(1991)}]{davis1991observing}%
  \BibitemOpen
  \bibfield  {author} {\bibinfo {author} {\bibfnamefont {R.~E.}\ \bibnamefont
  {Davis}},\ }\bibfield  {title} {\bibinfo {title} {Observing the general
  circulation with floats},\ }\href@noop {} {\bibfield  {journal} {\bibinfo
  {journal} {Deep Sea Research Part A. Oceanographic Research Papers}\ }\textbf
  {\bibinfo {volume} {38}},\ \bibinfo {pages} {S531} (\bibinfo {year}
  {1991})}\BibitemShut {NoStop}%
\bibitem [{\citenamefont {Fujii}\ \emph {et~al.}(2019)\citenamefont {Fujii},
  \citenamefont {R{\'e}my}, \citenamefont {Zuo}, \citenamefont {Oke},
  \citenamefont {Halliwell}, \citenamefont {Gasparin}, \citenamefont
  {Benkiran}, \citenamefont {Loose}, \citenamefont {Cummings}, \citenamefont
  {Xie} \emph {et~al.}}]{fujii2019observing}%
  \BibitemOpen
  \bibfield  {author} {\bibinfo {author} {\bibfnamefont {Y.}~\bibnamefont
  {Fujii}}, \bibinfo {author} {\bibfnamefont {E.}~\bibnamefont {R{\'e}my}},
  \bibinfo {author} {\bibfnamefont {H.}~\bibnamefont {Zuo}}, \bibinfo {author}
  {\bibfnamefont {P.}~\bibnamefont {Oke}}, \bibinfo {author} {\bibfnamefont
  {G.}~\bibnamefont {Halliwell}}, \bibinfo {author} {\bibfnamefont
  {F.}~\bibnamefont {Gasparin}}, \bibinfo {author} {\bibfnamefont
  {M.}~\bibnamefont {Benkiran}}, \bibinfo {author} {\bibfnamefont
  {N.}~\bibnamefont {Loose}}, \bibinfo {author} {\bibfnamefont
  {J.}~\bibnamefont {Cummings}}, \bibinfo {author} {\bibfnamefont
  {J.}~\bibnamefont {Xie}}, \emph {et~al.},\ }\bibfield  {title} {\bibinfo
  {title} {Observing system evaluation based on ocean data assimilation and
  prediction systems: on-going challenges and a future vision for designing and
  supporting ocean observational networks},\ }\href@noop {} {\bibfield
  {journal} {\bibinfo  {journal} {Frontiers in Marine Science}\ }\textbf
  {\bibinfo {volume} {6}},\ \bibinfo {pages} {417} (\bibinfo {year}
  {2019})}\BibitemShut {NoStop}%
\bibitem [{\citenamefont {Schlueter-Kuck}\ and\ \citenamefont
  {Dabiri}(2017{\natexlab{a}})}]{schlueter2017coherent}%
  \BibitemOpen
  \bibfield  {author} {\bibinfo {author} {\bibfnamefont {K.~L.}\ \bibnamefont
  {Schlueter-Kuck}}\ and\ \bibinfo {author} {\bibfnamefont {J.~O.}\
  \bibnamefont {Dabiri}},\ }\bibfield  {title} {\bibinfo {title} {Coherent
  structure colouring: identification of coherent structures from sparse data
  using graph theory},\ }\href@noop {} {\bibfield  {journal} {\bibinfo
  {journal} {Journal of Fluid Mechanics}\ }\textbf {\bibinfo {volume} {811}},\
  \bibinfo {pages} {468} (\bibinfo {year} {2017}{\natexlab{a}})}\BibitemShut
  {NoStop}%
\bibitem [{\citenamefont {Schlueter-Kuck}\ and\ \citenamefont
  {Dabiri}(2017{\natexlab{b}})}]{schlueter2017identification}%
  \BibitemOpen
  \bibfield  {author} {\bibinfo {author} {\bibfnamefont {K.~L.}\ \bibnamefont
  {Schlueter-Kuck}}\ and\ \bibinfo {author} {\bibfnamefont {J.~O.}\
  \bibnamefont {Dabiri}},\ }\bibfield  {title} {\bibinfo {title}
  {Identification of individual coherent sets associated with flow trajectories
  using coherent structure coloring},\ }\href@noop {} {\bibfield  {journal}
  {\bibinfo  {journal} {Chaos: An Interdisciplinary Journal of Nonlinear
  Science}\ }\textbf {\bibinfo {volume} {27}},\ \bibinfo {pages} {091101}
  (\bibinfo {year} {2017}{\natexlab{b}})}\BibitemShut {NoStop}%
\bibitem [{\citenamefont {Schlueter-Kuck}\ and\ \citenamefont
  {Dabiri}(2019)}]{schlueter2019model}%
  \BibitemOpen
  \bibfield  {author} {\bibinfo {author} {\bibfnamefont {K.~L.}\ \bibnamefont
  {Schlueter-Kuck}}\ and\ \bibinfo {author} {\bibfnamefont {J.~O.}\
  \bibnamefont {Dabiri}},\ }\bibfield  {title} {\bibinfo {title} {Model
  parameter estimation using coherent structure colouring},\ }\href@noop {}
  {\bibfield  {journal} {\bibinfo  {journal} {Journal of Fluid Mechanics}\
  }\textbf {\bibinfo {volume} {861}},\ \bibinfo {pages} {886} (\bibinfo {year}
  {2019})}\BibitemShut {NoStop}%
\bibitem [{\citenamefont {Husic}\ \emph {et~al.}(2019)\citenamefont {Husic},
  \citenamefont {Schlueter-Kuck},\ and\ \citenamefont
  {Dabiri}}]{husic2019simultaneous}%
  \BibitemOpen
  \bibfield  {author} {\bibinfo {author} {\bibfnamefont {B.~E.}\ \bibnamefont
  {Husic}}, \bibinfo {author} {\bibfnamefont {K.~L.}\ \bibnamefont
  {Schlueter-Kuck}},\ and\ \bibinfo {author} {\bibfnamefont {J.~O.}\
  \bibnamefont {Dabiri}},\ }\bibfield  {title} {\bibinfo {title} {Simultaneous
  coherent structure coloring facilitates interpretable clustering of
  scientific data by amplifying dissimilarity},\ }\href@noop {} {\bibfield
  {journal} {\bibinfo  {journal} {Plos one}\ }\textbf {\bibinfo {volume}
  {14}},\ \bibinfo {pages} {e0212442} (\bibinfo {year} {2019})}\BibitemShut
  {NoStop}%
\bibitem [{\citenamefont {Saqib}\ \emph {et~al.}(2017)\citenamefont {Saqib},
  \citenamefont {Khan},\ and\ \citenamefont
  {Blumenstein}}]{saqib2017detecting}%
  \BibitemOpen
  \bibfield  {author} {\bibinfo {author} {\bibfnamefont {M.}~\bibnamefont
  {Saqib}}, \bibinfo {author} {\bibfnamefont {S.~D.}\ \bibnamefont {Khan}},\
  and\ \bibinfo {author} {\bibfnamefont {M.}~\bibnamefont {Blumenstein}},\
  }\bibfield  {title} {\bibinfo {title} {Detecting dominant motion patterns in
  crowds of pedestrians},\ }in\ \href@noop {} {\emph {\bibinfo {booktitle}
  {Eighth International Conference on Graphic and Image Processing (ICGIP
  2016)}}},\ Vol.\ \bibinfo {volume} {10225}\ (\bibinfo {organization}
  {International Society for Optics and Photonics},\ \bibinfo {year} {2017})\
  p.\ \bibinfo {pages} {102251L}\BibitemShut {NoStop}%
\bibitem [{\citenamefont {Zhao}\ \emph {et~al.}(2020)\citenamefont {Zhao},
  \citenamefont {Hung},\ and\ \citenamefont {Wu}}]{zhao2020k}%
  \BibitemOpen
  \bibfield  {author} {\bibinfo {author} {\bibfnamefont {F.}~\bibnamefont
  {Zhao}}, \bibinfo {author} {\bibfnamefont {D.~L.}\ \bibnamefont {Hung}},\
  and\ \bibinfo {author} {\bibfnamefont {S.}~\bibnamefont {Wu}},\ }\bibfield
  {title} {\bibinfo {title} {K-means clustering-driven detection of
  time-resolved vortex patterns and cyclic variations inside a direct injection
  engine},\ }\href@noop {} {\bibfield  {journal} {\bibinfo  {journal} {Applied
  Thermal Engineering}\ }\textbf {\bibinfo {volume} {180}},\ \bibinfo {pages}
  {115810} (\bibinfo {year} {2020})}\BibitemShut {NoStop}%
\bibitem [{\citenamefont {Hadjighasem}\ \emph {et~al.}(2016)\citenamefont
  {Hadjighasem}, \citenamefont {Karrasch}, \citenamefont {Teramoto},\ and\
  \citenamefont {Haller}}]{hadjighasem2016spectral}%
  \BibitemOpen
  \bibfield  {author} {\bibinfo {author} {\bibfnamefont {A.}~\bibnamefont
  {Hadjighasem}}, \bibinfo {author} {\bibfnamefont {D.}~\bibnamefont
  {Karrasch}}, \bibinfo {author} {\bibfnamefont {H.}~\bibnamefont {Teramoto}},\
  and\ \bibinfo {author} {\bibfnamefont {G.}~\bibnamefont {Haller}},\
  }\bibfield  {title} {\bibinfo {title} {Spectral-clustering approach to
  lagrangian vortex detection},\ }\href@noop {} {\bibfield  {journal} {\bibinfo
   {journal} {Physical Review E}\ }\textbf {\bibinfo {volume} {93}},\ \bibinfo
  {pages} {063107} (\bibinfo {year} {2016})}\BibitemShut {NoStop}%
\bibitem [{\citenamefont {Martins}\ \emph {et~al.}(2021)\citenamefont
  {Martins}, \citenamefont {Sciacchitano},\ and\ \citenamefont
  {Rival}}]{martins2021detection}%
  \BibitemOpen
  \bibfield  {author} {\bibinfo {author} {\bibfnamefont {F.}~\bibnamefont
  {Martins}}, \bibinfo {author} {\bibfnamefont {A.}~\bibnamefont
  {Sciacchitano}},\ and\ \bibinfo {author} {\bibfnamefont {D.}~\bibnamefont
  {Rival}},\ }\bibfield  {title} {\bibinfo {title} {Detection of vortical
  structures in sparse lagrangian data using coherent-structure colouring},\
  }\href@noop {} {\bibfield  {journal} {\bibinfo  {journal} {Experiments in
  Fluids}\ }\textbf {\bibinfo {volume} {62}},\ \bibinfo {pages} {1} (\bibinfo
  {year} {2021})}\BibitemShut {NoStop}%
\bibitem [{\citenamefont {DiMarco}\ \emph {et~al.}(2005)\citenamefont
  {DiMarco}, \citenamefont {Nowlin},\ and\ \citenamefont
  {Reid}}]{dimarco2005statistical}%
  \BibitemOpen
  \bibfield  {author} {\bibinfo {author} {\bibfnamefont {S.~F.}\ \bibnamefont
  {DiMarco}}, \bibinfo {author} {\bibfnamefont {W.~D.}\ \bibnamefont
  {Nowlin}},\ and\ \bibinfo {author} {\bibfnamefont {R.}~\bibnamefont {Reid}},\
  }\bibfield  {title} {\bibinfo {title} {A statistical description of the
  velocity fields from upper ocean drifters in the gulf of mexico},\
  }\href@noop {} {\bibfield  {journal} {\bibinfo  {journal} {Geophysical
  Monograph-American Geophysical Union}\ }\textbf {\bibinfo {volume} {161}},\
  \bibinfo {pages} {101} (\bibinfo {year} {2005})}\BibitemShut {NoStop}%
\bibitem [{\citenamefont {Espanol}\ and\ \citenamefont
  {Serrano}(2009)}]{espanol2009voronoi}%
  \BibitemOpen
  \bibfield  {author} {\bibinfo {author} {\bibfnamefont {P.}~\bibnamefont
  {Espanol}}\ and\ \bibinfo {author} {\bibfnamefont {M.}~\bibnamefont
  {Serrano}},\ }\bibfield  {title} {\bibinfo {title} {Voronoi fluid particles
  \& tessellation fluid dynamics},\ }\href@noop {} {\bibfield  {journal}
  {\bibinfo  {journal} {Tessellations in the Sciences Virtues, Techniques and
  Applications of Geometric Tilings}\ }\textbf {\bibinfo {volume} {3}},\
  \bibinfo {pages} {42} (\bibinfo {year} {2009})}\BibitemShut {NoStop}%
\bibitem [{\citenamefont {Rosi}\ \emph {et~al.}(2015)\citenamefont {Rosi},
  \citenamefont {Walker},\ and\ \citenamefont {Rival}}]{rosi2015lagrangian}%
  \BibitemOpen
  \bibfield  {author} {\bibinfo {author} {\bibfnamefont {G.~A.}\ \bibnamefont
  {Rosi}}, \bibinfo {author} {\bibfnamefont {A.~M.}\ \bibnamefont {Walker}},\
  and\ \bibinfo {author} {\bibfnamefont {D.~E.}\ \bibnamefont {Rival}},\
  }\bibfield  {title} {\bibinfo {title} {Lagrangian coherent structure
  identification using a voronoi tessellation-based networking algorithm},\
  }\href@noop {} {\bibfield  {journal} {\bibinfo  {journal} {Experiments in
  Fluids}\ }\textbf {\bibinfo {volume} {56}},\ \bibinfo {pages} {189} (\bibinfo
  {year} {2015})}\BibitemShut {NoStop}%
\bibitem [{\citenamefont {Krueger}\ \emph {et~al.}(2019)\citenamefont
  {Krueger}, \citenamefont {Hahsler}, \citenamefont {Olinick}, \citenamefont
  {Williams},\ and\ \citenamefont {Zharfa}}]{krueger2019quantitative}%
  \BibitemOpen
  \bibfield  {author} {\bibinfo {author} {\bibfnamefont {P.~S.}\ \bibnamefont
  {Krueger}}, \bibinfo {author} {\bibfnamefont {M.}~\bibnamefont {Hahsler}},
  \bibinfo {author} {\bibfnamefont {E.~V.}\ \bibnamefont {Olinick}}, \bibinfo
  {author} {\bibfnamefont {S.~H.}\ \bibnamefont {Williams}},\ and\ \bibinfo
  {author} {\bibfnamefont {M.}~\bibnamefont {Zharfa}},\ }\bibfield  {title}
  {\bibinfo {title} {Quantitative classification of vortical flows based on
  topological features using graph matching},\ }\href@noop {} {\bibfield
  {journal} {\bibinfo  {journal} {Proceedings of the Royal Society A}\ }\textbf
  {\bibinfo {volume} {475}},\ \bibinfo {pages} {20180897} (\bibinfo {year}
  {2019})}\BibitemShut {NoStop}%
\bibitem [{\citenamefont {Neeteson}\ and\ \citenamefont
  {Rival}(2015)}]{neeteson2015pressure}%
  \BibitemOpen
  \bibfield  {author} {\bibinfo {author} {\bibfnamefont {N.~J.}\ \bibnamefont
  {Neeteson}}\ and\ \bibinfo {author} {\bibfnamefont {D.~E.}\ \bibnamefont
  {Rival}},\ }\bibfield  {title} {\bibinfo {title} {Pressure-field extraction
  on unstructured flow data using a voronoi tessellation-based networking
  algorithm: a proof-of-principle study},\ }\href@noop {} {\bibfield  {journal}
  {\bibinfo  {journal} {Experiments in Fluids}\ }\textbf {\bibinfo {volume}
  {56}},\ \bibinfo {pages} {44} (\bibinfo {year} {2015})}\BibitemShut {NoStop}%
\bibitem [{\citenamefont {Ferrero}(2011)}]{ferrero2011voronoi}%
  \BibitemOpen
  \bibfield  {author} {\bibinfo {author} {\bibfnamefont {M.}~\bibnamefont
  {Ferrero}},\ }\bibfield  {title} {\bibinfo {title} {Voronoi diagram: The
  generator recognition problem},\ }\href@noop {} {\bibfield  {journal}
  {\bibinfo  {journal} {arXiv preprint arXiv:1105.4246}\ } (\bibinfo {year}
  {2011})}\BibitemShut {NoStop}%
\bibitem [{\citenamefont {Chew}(1989)}]{chew1989constrained}%
  \BibitemOpen
  \bibfield  {author} {\bibinfo {author} {\bibfnamefont {L.~P.}\ \bibnamefont
  {Chew}},\ }\bibfield  {title} {\bibinfo {title} {Constrained delaunay
  triangulations},\ }\href@noop {} {\bibfield  {journal} {\bibinfo  {journal}
  {Algorithmica}\ }\textbf {\bibinfo {volume} {4}},\ \bibinfo {pages} {97}
  (\bibinfo {year} {1989})}\BibitemShut {NoStop}%
\bibitem [{\citenamefont {Spielman}(2007)}]{spielman2007spectral}%
  \BibitemOpen
  \bibfield  {author} {\bibinfo {author} {\bibfnamefont {D.~A.}\ \bibnamefont
  {Spielman}},\ }\href@noop {} {\emph {\bibinfo {title} {48th Annual IEEE
  Symposium on Foundations of Computer Science (FOCS'07)}}}\ (\bibinfo
  {publisher} {IEEE},\ \bibinfo {year} {2007})\ pp.\ \bibinfo {pages}
  {29--38}\BibitemShut {NoStop}%
\bibitem [{\citenamefont {Haller}(2000)}]{haller2000finding}%
  \BibitemOpen
  \bibfield  {author} {\bibinfo {author} {\bibfnamefont {G.}~\bibnamefont
  {Haller}},\ }\bibfield  {title} {\bibinfo {title} {Finding finite-time
  invariant manifolds in two-dimensional velocity fields},\ }\href@noop {}
  {\bibfield  {journal} {\bibinfo  {journal} {Chaos: An Interdisciplinary
  Journal of Nonlinear Science}\ }\textbf {\bibinfo {volume} {10}},\ \bibinfo
  {pages} {99} (\bibinfo {year} {2000})}\BibitemShut {NoStop}%
\bibitem [{\citenamefont {Shadden}\ \emph {et~al.}(2005)\citenamefont
  {Shadden}, \citenamefont {Lekien},\ and\ \citenamefont
  {Marsden}}]{shadden2005definition}%
  \BibitemOpen
  \bibfield  {author} {\bibinfo {author} {\bibfnamefont {S.~C.}\ \bibnamefont
  {Shadden}}, \bibinfo {author} {\bibfnamefont {F.}~\bibnamefont {Lekien}},\
  and\ \bibinfo {author} {\bibfnamefont {J.~E.}\ \bibnamefont {Marsden}},\
  }\bibfield  {title} {\bibinfo {title} {Definition and properties of
  lagrangian coherent structures from finite-time lyapunov exponents in
  two-dimensional aperiodic flows},\ }\href@noop {} {\bibfield  {journal}
  {\bibinfo  {journal} {Physica D: Nonlinear Phenomena}\ }\textbf {\bibinfo
  {volume} {212}},\ \bibinfo {pages} {271} (\bibinfo {year}
  {2005})}\BibitemShut {NoStop}%
\bibitem [{\citenamefont {Senatore}\ and\ \citenamefont
  {Ross}(2011)}]{senatore2011detection}%
  \BibitemOpen
  \bibfield  {author} {\bibinfo {author} {\bibfnamefont {C.}~\bibnamefont
  {Senatore}}\ and\ \bibinfo {author} {\bibfnamefont {S.~D.}\ \bibnamefont
  {Ross}},\ }\bibfield  {title} {\bibinfo {title} {Detection and
  characterization of transport barriers in complex flows via ridge extraction
  of the finite time lyapunov exponent field},\ }\href@noop {} {\bibfield
  {journal} {\bibinfo  {journal} {International Journal for Numerical Methods
  in Engineering}\ }\textbf {\bibinfo {volume} {86}},\ \bibinfo {pages} {1163}
  (\bibinfo {year} {2011})}\BibitemShut {NoStop}%
\bibitem [{\citenamefont {Williams}\ \emph {et~al.}(2015)\citenamefont
  {Williams}, \citenamefont {Rypina},\ and\ \citenamefont
  {Rowley}}]{williams2015identifying}%
  \BibitemOpen
  \bibfield  {author} {\bibinfo {author} {\bibfnamefont {M.~O.}\ \bibnamefont
  {Williams}}, \bibinfo {author} {\bibfnamefont {I.~I.}\ \bibnamefont
  {Rypina}},\ and\ \bibinfo {author} {\bibfnamefont {C.~W.}\ \bibnamefont
  {Rowley}},\ }\bibfield  {title} {\bibinfo {title} {Identifying finite-time
  coherent sets from limited quantities of lagrangian data},\ }\href@noop {}
  {\bibfield  {journal} {\bibinfo  {journal} {Chaos: An Interdisciplinary
  Journal of Nonlinear Science}\ }\textbf {\bibinfo {volume} {25}},\ \bibinfo
  {pages} {087408} (\bibinfo {year} {2015})}\BibitemShut {NoStop}%
\bibitem [{\citenamefont {Pratt}\ \emph {et~al.}(2015)\citenamefont {Pratt},
  \citenamefont {Meiss},\ and\ \citenamefont {Crimaldi}}]{pratt2015reaction}%
  \BibitemOpen
  \bibfield  {author} {\bibinfo {author} {\bibfnamefont {K.~R.}\ \bibnamefont
  {Pratt}}, \bibinfo {author} {\bibfnamefont {J.~D.}\ \bibnamefont {Meiss}},\
  and\ \bibinfo {author} {\bibfnamefont {J.~P.}\ \bibnamefont {Crimaldi}},\
  }\bibfield  {title} {\bibinfo {title} {Reaction enhancement of initially
  distant scalars by lagrangian coherent structures},\ }\href@noop {}
  {\bibfield  {journal} {\bibinfo  {journal} {Physics of Fluids}\ }\textbf
  {\bibinfo {volume} {27}},\ \bibinfo {pages} {035106} (\bibinfo {year}
  {2015})}\BibitemShut {NoStop}%
\bibitem [{\citenamefont {Balasuriya}(2016)}]{balasuriya2016barriers}%
  \BibitemOpen
  \bibfield  {author} {\bibinfo {author} {\bibfnamefont {S.}~\bibnamefont
  {Balasuriya}},\ }\href@noop {} {\emph {\bibinfo {title} {Barriers and
  transport in unsteady flows: a Melnikov approach}}}\ (\bibinfo  {publisher}
  {SIAM},\ \bibinfo {year} {2016})\BibitemShut {NoStop}%
\bibitem [{\citenamefont {Balasuriya}\ \emph {et~al.}(2018)\citenamefont
  {Balasuriya}, \citenamefont {Ouellette},\ and\ \citenamefont
  {Rypina}}]{balasuriya2018generalized}%
  \BibitemOpen
  \bibfield  {author} {\bibinfo {author} {\bibfnamefont {S.}~\bibnamefont
  {Balasuriya}}, \bibinfo {author} {\bibfnamefont {N.~T.}\ \bibnamefont
  {Ouellette}},\ and\ \bibinfo {author} {\bibfnamefont {I.~I.}\ \bibnamefont
  {Rypina}},\ }\bibfield  {title} {\bibinfo {title} {Generalized lagrangian
  coherent structures},\ }\href@noop {} {\bibfield  {journal} {\bibinfo
  {journal} {Physica D: Nonlinear Phenomena}\ }\textbf {\bibinfo {volume}
  {372}},\ \bibinfo {pages} {31} (\bibinfo {year} {2018})}\BibitemShut
  {NoStop}%
\bibitem [{\citenamefont {Hou}\ \emph {et~al.}(2021)\citenamefont {Hou},
  \citenamefont {Kaiser}, \citenamefont {Sciacchitano},\ and\ \citenamefont
  {Rival}}]{hou2021novel}%
  \BibitemOpen
  \bibfield  {author} {\bibinfo {author} {\bibfnamefont {J.}~\bibnamefont
  {Hou}}, \bibinfo {author} {\bibfnamefont {F.}~\bibnamefont {Kaiser}},
  \bibinfo {author} {\bibfnamefont {A.}~\bibnamefont {Sciacchitano}},\ and\
  \bibinfo {author} {\bibfnamefont {D.~E.}\ \bibnamefont {Rival}},\ }\bibfield
  {title} {\bibinfo {title} {A novel single-camera approach to large-scale,
  three-dimensional particle tracking based on glare-point spacing},\
  }\href@noop {} {\bibfield  {journal} {\bibinfo  {journal} {Experiments in
  Fluids}\ }\textbf {\bibinfo {volume} {62}},\ \bibinfo {pages} {1} (\bibinfo
  {year} {2021})}\BibitemShut {NoStop}%
\bibitem [{\citenamefont {Schneiders}\ and\ \citenamefont
  {Scarano}(2016)}]{schneiders2016dense}%
  \BibitemOpen
  \bibfield  {author} {\bibinfo {author} {\bibfnamefont {J.~F.}\ \bibnamefont
  {Schneiders}}\ and\ \bibinfo {author} {\bibfnamefont {F.}~\bibnamefont
  {Scarano}},\ }\bibfield  {title} {\bibinfo {title} {Dense velocity
  reconstruction from tomographic ptv with material derivatives},\ }\href@noop
  {} {\bibfield  {journal} {\bibinfo  {journal} {Experiments in fluids}\
  }\textbf {\bibinfo {volume} {57}},\ \bibinfo {pages} {1} (\bibinfo {year}
  {2016})}\BibitemShut {NoStop}%
\bibitem [{\citenamefont {Peng}\ and\ \citenamefont
  {Dabiri}(2009)}]{peng2009transport}%
  \BibitemOpen
  \bibfield  {author} {\bibinfo {author} {\bibfnamefont {J.}~\bibnamefont
  {Peng}}\ and\ \bibinfo {author} {\bibfnamefont {J.}~\bibnamefont {Dabiri}},\
  }\bibfield  {title} {\bibinfo {title} {Transport of inertial particles by
  lagrangian coherent structures: application to predator--prey interaction in
  jellyfish feeding},\ }\href@noop {} {\bibfield  {journal} {\bibinfo
  {journal} {Journal of Fluid Mechanics}\ }\textbf {\bibinfo {volume} {623}},\
  \bibinfo {pages} {75} (\bibinfo {year} {2009})}\BibitemShut {NoStop}%
\bibitem [{\citenamefont {Garcia}(2010)}]{garcia2010robust}%
  \BibitemOpen
  \bibfield  {author} {\bibinfo {author} {\bibfnamefont {D.}~\bibnamefont
  {Garcia}},\ }\bibfield  {title} {\bibinfo {title} {Robust smoothing of
  gridded data in one and higher dimensions with missing values},\ }\href@noop
  {} {\bibfield  {journal} {\bibinfo  {journal} {Computational statistics \&
  data analysis}\ }\textbf {\bibinfo {volume} {54}},\ \bibinfo {pages} {1167}
  (\bibinfo {year} {2010})}\BibitemShut {NoStop}%
\bibitem [{\citenamefont {Garth}\ and\ \citenamefont
  {Joy}(2010)}]{garth2010fast}%
  \BibitemOpen
  \bibfield  {author} {\bibinfo {author} {\bibfnamefont {C.}~\bibnamefont
  {Garth}}\ and\ \bibinfo {author} {\bibfnamefont {K.~I.}\ \bibnamefont
  {Joy}},\ }\bibfield  {title} {\bibinfo {title} {Fast, memory-efficient cell
  location in unstructured grids for visualization},\ }\href@noop {} {\bibfield
   {journal} {\bibinfo  {journal} {IEEE Transactions on Visualization and
  Computer Graphics}\ }\textbf {\bibinfo {volume} {16}},\ \bibinfo {pages}
  {1541} (\bibinfo {year} {2010})}\BibitemShut {NoStop}%
\bibitem [{\citenamefont {Moukalled}\ \emph {et~al.}(2016)\citenamefont
  {Moukalled}, \citenamefont {Mangani}, \citenamefont {Darwish} \emph
  {et~al.}}]{moukalled2016finite}%
  \BibitemOpen
  \bibfield  {author} {\bibinfo {author} {\bibfnamefont {F.}~\bibnamefont
  {Moukalled}}, \bibinfo {author} {\bibfnamefont {L.}~\bibnamefont {Mangani}},
  \bibinfo {author} {\bibfnamefont {M.}~\bibnamefont {Darwish}}, \emph
  {et~al.},\ }\href@noop {} {\emph {\bibinfo {title} {The finite volume method
  in computational fluid dynamics}}},\ Vol.~\bibinfo {volume} {6}\ (\bibinfo
  {publisher} {Springer},\ \bibinfo {year} {2016})\BibitemShut {NoStop}%
\bibitem [{\citenamefont {Henn}\ \emph {et~al.}(2013)\citenamefont {Henn},
  \citenamefont {Raleigh}, \citenamefont {Fisher},\ and\ \citenamefont
  {Lundquist}}]{henn2013comparison}%
  \BibitemOpen
  \bibfield  {author} {\bibinfo {author} {\bibfnamefont {B.}~\bibnamefont
  {Henn}}, \bibinfo {author} {\bibfnamefont {M.~S.}\ \bibnamefont {Raleigh}},
  \bibinfo {author} {\bibfnamefont {A.}~\bibnamefont {Fisher}},\ and\ \bibinfo
  {author} {\bibfnamefont {J.~D.}\ \bibnamefont {Lundquist}},\ }\bibfield
  {title} {\bibinfo {title} {A comparison of methods for filling gaps in hourly
  near-surface air temperature data},\ }\href@noop {} {\bibfield  {journal}
  {\bibinfo  {journal} {Journal of Hydrometeorology}\ }\textbf {\bibinfo
  {volume} {14}},\ \bibinfo {pages} {929} (\bibinfo {year} {2013})}\BibitemShut
  {NoStop}%
\bibitem [{\citenamefont {Marshall}(2005)}]{marshall2005particle}%
  \BibitemOpen
  \bibfield  {author} {\bibinfo {author} {\bibfnamefont {J.}~\bibnamefont
  {Marshall}},\ }\bibfield  {title} {\bibinfo {title} {Particle dispersion in a
  turbulent vortex core},\ }\href@noop {} {\bibfield  {journal} {\bibinfo
  {journal} {Physics of Fluids}\ }\textbf {\bibinfo {volume} {17}},\ \bibinfo
  {pages} {025104} (\bibinfo {year} {2005})}\BibitemShut {NoStop}%
\bibitem [{\citenamefont {Faleiros}\ \emph {et~al.}(2019)\citenamefont
  {Faleiros}, \citenamefont {Tuinstra}, \citenamefont {Sciacchitano},\ and\
  \citenamefont {Scarano}}]{faleiros2019generation}%
  \BibitemOpen
  \bibfield  {author} {\bibinfo {author} {\bibfnamefont {D.~E.}\ \bibnamefont
  {Faleiros}}, \bibinfo {author} {\bibfnamefont {M.}~\bibnamefont {Tuinstra}},
  \bibinfo {author} {\bibfnamefont {A.}~\bibnamefont {Sciacchitano}},\ and\
  \bibinfo {author} {\bibfnamefont {F.}~\bibnamefont {Scarano}},\ }\bibfield
  {title} {\bibinfo {title} {Generation and control of helium-filled soap
  bubbles for piv},\ }\href@noop {} {\bibfield  {journal} {\bibinfo  {journal}
  {Experiments in Fluids}\ }\textbf {\bibinfo {volume} {60}},\ \bibinfo {pages}
  {1} (\bibinfo {year} {2019})}\BibitemShut {NoStop}%
\bibitem [{\citenamefont {Brunton}\ and\ \citenamefont
  {Rowley}(2010)}]{brunton2010fast}%
  \BibitemOpen
  \bibfield  {author} {\bibinfo {author} {\bibfnamefont {S.~L.}\ \bibnamefont
  {Brunton}}\ and\ \bibinfo {author} {\bibfnamefont {C.~W.}\ \bibnamefont
  {Rowley}},\ }\bibfield  {title} {\bibinfo {title} {Fast computation of ftle
  fields for unsteady flows: a comparison of methods},\ }\href@noop {}
  {\bibfield  {journal} {\bibinfo  {journal} {Chaos}\ }\textbf {\bibinfo
  {volume} {20}},\ \bibinfo {pages} {017503} (\bibinfo {year}
  {2010})}\BibitemShut {NoStop}%
\end{thebibliography}%

\end{document}